\documentclass[reprint,superscriptaddress,preprintnumbers,nofootinbib,amsmath,amssymb,aps,prd,showkeys,showpacs]{revtex4-2}

\usepackage{graphicx}% Include figure files
\usepackage{dcolumn}% Align table columns on decimal point
\usepackage{bm}% bold math
\usepackage{array} % 导言区引入
\newcolumntype{M}[1]{>{\centering\arraybackslash}m{#1}}
\usepackage{endnotes} 

\usepackage{amssymb}    %Allows use math symbols
\usepackage{color}         %Allows {\color{red} Hello World} or \colorbox{blue}{Hello World}
\usepackage{graphicx}     %Allows\includegraphics{figure.pdf}
\usepackage{mathrsfs} %Allows \mathscr{HELLO}
\usepackage{hyperref} %Automatically links \label and \ref commands; Always load last
\hypersetup{colorlinks=true,linkcolor=blue,citecolor=blue}
\usepackage{ upgreek } % for uptau
\usepackage{color,xcolor,fancybox,epsf,rotating,colordvi}
\usepackage{booktabs}
\usepackage{multirow}

\def\add#1{\textcolor{blue}{#1}}

\newcommand{\SLASH}[2]{\makebox[#2ex][l]{$#1$}/}

\newcommand{\tanb}{\tan\!\beta}

\newcommand{\GeV}{~\rm GeV}

\newcommand{\fbm}{{~\rm fb}^{-1}}

\newcommand{\abm}{{~\rm ab}^{-1}}
\begin{document}

% \preprint{APS/123-QED}

\title{Probing a Type-I 2HDM light Higgs boson in the top-pair-associated diphoton channel}

\author{Yabo Dong}
\email[]{dongyb@henu.edu.cn}
\affiliation{School of Physics and Electronics, Henan University, Kaifeng 475004, China}

\author{Kun Wang}
% \email[]{kwang@usst.edu.cn}
\email[Corresponding author:]{kwang@usst.edu.cn} 
\affiliation{College of Science, University of Shanghai for Science and Technology, Shanghai 200093, China}

\author{Jingya Zhu}
% \email[]{zhujy@henu.edu.cn} 
\email[Corresponding author:]{zhujy@henu.edu.cn} 
\affiliation{School of Physics and Electronics, Henan University, Kaifeng 475004, China}

% \collaboration{CLEO Collaboration}%\noaffiliation

\date{September 9, 2025}

\begin{abstract}
Motivated by the possible 95 GeV diphoton excess, we investigate the capability of the Type-I Two-Higgs-Doublet Model (2HDM-I) to explain this signal under current theoretical and experimental constraints. Using full Monte Carlo (MC) simulations for the process of $pp \to t(\to W^+ b)\bar{t}(\to W^- \bar{b})h(\to \gamma\gamma)$, we evaluate the discovery potential of a 95 GeV Higgs boson at future colliders. Direct Higgs searches strongly constrain the parameter $\alpha$, excluding the region with $\alpha \lesssim 0.95$. 
Monte Carlo results indicate that a minimum cross section of 0.3 fb is required to achieve a $5\sigma$ signal statistical significance at the HL-LHC with $L = 3~\mathrm{ab}^{-1}$. 
For the same luminosity, HE-LHC and FCC-hh require 0.67 fb and 2.36 fb, respectively. 
At the 14 TeV HL-LHC with an integrated luminosity of $3~\mathrm{ab}^{-1}$, parameter regions with $\sin(\beta-\alpha) \gtrsim 0.4$ and $\sin(\beta-\alpha) \gtrsim 0.25$ can be probed at the $5\sigma$ and $2\sigma$ significance levels, respectively. At the 27 TeV HE-LHC with $L = 10~\mathrm{ab}^{-1}$, the sensitivity improves to $\sin(\beta-\alpha) \gtrsim 0.25$ ($5\sigma$) and $\gtrsim 0.15$ ($2\sigma$). For the 100 TeV FCC-hh with $L = 30~\mathrm{ab}^{-1}$, even regions with $\sin(\beta-\alpha) \gtrsim 0.1$ or $\sin(\beta-\alpha) \lesssim -0.05$ can be covered at the $5\sigma$ level. 
Parameter regions near $\sin(\beta-\alpha) \approx 0$ remain challenging to probe in the diphoton channel, even with increased energy or luminosity. 
\end{abstract}

\maketitle
\newpage

% \tableofcontents
% \newpage

%%%

\section{Introduction}
\label{sec:intro}
The primary goal of the LHC is to search for the Higgs boson and hypothetical particles beyond the Standard Model (BSM). After the decisive discovery of the 125 GeV Higgs boson by the CMS \cite{CMS:2012qbp} and ATLAS \cite{ATLAS:2012yve} collaborations in 2012, subsequent measurements have confirmed its mass, spin, width, and couplings \cite{CMS:2022dwd, ATLAS:2022vkf}, largely confirming the predictions of the Standard Model (SM). 
While the SM remains robust under these experimental validations, the absence of new particle discoveries significantly restricts the parameter space of BSM theories. 
Nevertheless, persistent anomalies and excess signals observed in experiments at LEP, Tevatron, and LHC hint at the presence of new physics, underscoring the importance of continued investigations into BSM theories. 
These studies are crucial for defining the limits of the SM and guiding future theoretical and experimental efforts.

As early as 2003, the LEP experiment documented a 2.3$\sigma$ local excess in the $e^+e^-\rightarrow Z(h\rightarrow b\bar{b})$ channel at about 98 GeV \cite{LEPWorkingGroupforHiggsbosonsearches:2003ing} 
(here and throughout this paper, $h$ denotes the hypothetical lighter scalar with a mass of 95 GeV; the observed 125 GeV Higgs boson is denoted by $H$). 
In 2018, the CMS collaboration combined data from 8 TeV (19.7 $\fbm$) and 13 TeV (35.9 $\fbm$) runs to report a $h\rightarrow \gamma \gamma$ excess for a mass of 95.3 GeV with the local (global) significance of 2.8$\sigma$ (1.3$\sigma$), respectively \cite{CMS:2018cyk}. 
More recently, in 2023, ATLAS reported a 1.7$\sigma$ excess at a mass of 95.4 GeV in the $h\rightarrow \gamma \gamma$ channel \cite{ATLAS:2023jzc}. 
In 2024, CMS reported a local (global) significance of 2.9$\sigma$ (1.3$\sigma$) for the same mass hypothesis \cite{CMS:2024yhz}. 
At the closely related diphoton invariant mass points of 95.4 GeV and 95.3 GeV, both ATLAS and CMS have reported a persistent signal excess. 
This excess has shown no sign of diminishing over time, strongly suggesting the possible existence of a scalar particle around 95 GeV in the diphoton channel.

The reported 95 GeV excess has attracted considerable theoretical interest as a possible signal of new physics \cite{Azevedo:2023zkg, Khanna:2024bah, Benbrik:2024ptw, Belyaev:2023xnv, Benbrik:2022azi, Ashanujjaman:2023etj, Ge:2024rdr, Ahriche:2023hho, Dev:2023kzu}. Recent investigations have meticulously examined this phenomenon across various models, focusing on the presence of a light scalar. 
These models include the simple SM extensions \cite{Kundu:2019nqo, Aguilar-Saavedra:2020wrj, YaserAyazi:2024hpj}, 
the minimal dilaton model \cite{Liu:2018ryo, Wang:2024bkg}, 
the two Higgs doublet model (2HDM) with a singlet extensions \cite{Heinemeyer:2021msz, Biekotter:2021ovi,  Biekotter:2023jld, Biekotter:2023oen, Arcadi:2023smv}, 
the next-to-minimal two Higgs doublet model and its extensions \cite{Biekotter:2019kde, Biekotter:2021qbc, Banik:2023ecr, Aguilar-Saavedra:2023tql}, 
the Georgi-Machacek model \cite{Wang:2022okq, Chen:2023bqr, Ahriche:2023wkj}, 
the next-to-minimal supersymmetric model \cite{Cao:2016uwt, Cao:2019ofo, Hollik:2020plc, Li:2022etb, Domingo:2022pde, Cao:2023gkc, Li:2023kbf, Lian:2024smg, Cao:2024axg, Ellwanger:2023zjc, Ellwanger:2024vvs} and its semi-constrained version \cite{Wang:2018vxp}, etc. 
For the model of these studies, the light scalar is usually singlet-dominated, mixing in part with the doublet Higgs. On the contrary, less attention has been paid to the model with a doublet-dominated scalar.

In this study, we explore the 2HDM, an additional $SU(2)_L$ Higgs doublet extension of the SM that includes a second Higgs doublet \cite{Lee:1973iz, Branco:2011iw, Atwood:1996vj, Gunion:2002zf, Bhattacharyya:2015nca, Wang:2022yhm}, to give a correlation explanation of the doublet-dominated scalar. 
The primary rationale for introducing a second Higgs doublet is rooted in the insufficiency of a single doublet to both endow up-type and down-type quarks with mass and to eliminate anomalies inherent in supersymmetry \cite{Haber:1984rc}. 
The theoretical framework of 2HDM predicts a pseudoscalar Higgs boson, a pair of charged Higgs bosons, and two neutral scalar Higgs bosons. The pseudoscalar Higgs or one of the two neutral scalar Higgs can be a candidate to explain the 95 GeV excess in the experiments. 
Meanwhile, the inclusion of charged Higgs bosons and pseudoscalar in the 2HDM enriches the phenomenological landscape, offering diverse decay modes and branching ratios beyond those in the SM. 

However, the 2HDM is particularly affected by experimental constraints due to its potential role in mediating flavor-changing neutral currents (FCNCs). To mitigate these effects, several variants of the 2HDM have been proposed \cite{Pich:2009sp, Tuzon:2010vt}, including type-I (2HDM-I), type-II (2HDM-II), lepton-specific (2HDM-LS), and a flipped model (2HDM-Flipped), etc. 
Specifically, in the 2HDM-I \cite{Haber:1978jt, Hall:1981bc, Bhattacharyya:2015nca, Arhrib:2017wmo, Wang:2021pxc}, a unique characteristic is that only one scalar doublet $\Phi _2$ couple to quarks and leptons. 
According to its coupling characteristics, 2HDM-I has some interesting features. 
For example, 2HDM-I can predict a light neutral scalar $h$ and a charged scalar $H^{\pm}$ under similar theoretical and experimental constraints \add{\cite{Fox:2017uwr, Haisch:2017gql}}.
Meanwhile, the light $H^{\pm}$ can give an additional contribution to the process of $h\to\gamma\gamma$ through the loop effect, thereby increasing the decay of $h$ to diphoton and suppress other decays \cite{Posch:2010hx}. 
Consequently, the light neutral scalar $h$ in 2HDM-I may be an outstanding candidate to explain the 95 GeV diphoton excess and this work primarily explores such phenomenology of 2HDM-I.

Top-pair-associated channel can effectively reduce the SM QCD backgrounds by providing more final state tags. It is meaningful to study this channel, especially after the SM-like Higgs is confirmed in this channel \cite{ATLAS:2021qou, ATLAS:2024gth}.
Therefore, this work focuses on the top-pair-associated diphoton channel. 
In this research, we explore the parameter space of the 2HDM-I model, evaluating its viability under current constraints to accommodate a 95 GeV light Higgs boson.
Following this analysis, we perform Monte Carlo (MC) simulations to predict the feasibility of testing this model in future collider experiments, with a focus on the top-pair-associated diphoton channel.

The organization of this paper is as follows.
In Sec.~\ref{sec:scenario}, we provide a brief overview of the 2HDM-I model.
Sec.~\ref{sec:scan} shows the region of the parameter space scan and the discussion of the samples surviving condition and muon g-2.
In Sec.~\ref{sec:result}, we conduct collider simulations and analyze the results.
In Sec.~\ref{sec:conclusion}, we present our conclusions.
\section{\label{sec:scenario}The two-Higgs-doublet model}

The 2HDM is a widely accepted and straightforward extension of the SM. The most general scalar potential in this model includes 14 additional parameters.  Assuming CP conservation in the Higgs sector, it involves two complex doublets, $\Phi_1$ and $\Phi_2$, each with a hypercharge $Y = +1$. The scalar potential is given by \cite{Lee:1973iz, Branco:2011iw, Atwood:1996vj}:
\begin{widetext}
\begin{align}
    V(\Phi _1, \Phi _2)  = & \quad m_{11}^{2} \Phi_{1}^{\dagger}\Phi_{1}+m_{22}^{2} \Phi_{2}^{\dagger}\Phi_{2}-m_{12}^{2} \left(\Phi_{1}^{\dagger}\Phi_{2}+ \mathrm{h.c.} \right) \\ &+\frac{\lambda_{1}}{2} \left(\Phi_{1}^{\dagger}\Phi_{1}\right)^{2} \nonumber + \frac{\lambda_{2}}{2} \left(\Phi_{2}^{\dagger}\Phi_{2}\right)^{2}+\lambda_{3} \Phi_{1}^{\dagger}\Phi_{1} \Phi_{2}^{\dagger}\Phi_{2}+\lambda_{4} \Phi_{1}^{\dagger}\Phi_{2} \Phi_{2}^{\dagger}\Phi_{1}+\frac{\lambda_{5}}{2}\left[\left(\Phi_{1}^{\dagger}\Phi_{2}\right)^{2}+ \mathrm{h.c.}\right],
\end{align}
\end{widetext}
where $m_{11}^2$, $m_{12}^2$ , $m_{22}^2$, and $\lambda_i(i=1,2...5)$ are real parameters. 
The model contains two complex scalar SU(2) doublets, which are defined as
\begin{align}
\Phi_i=\left(\begin{array}{c}\phi_i^+ \\ (v_i+\phi_i+i\eta_i) \Big/\sqrt{2}\end{array}\right),\quad i=1,2\,,
\end{align}
where $v_i$ is the vacuum expectation values (VEVs) of $\Phi_i$, with $\sqrt{v_1^2 + v_2^2} \equiv v \approx 246 \GeV$.

The free parameters can then be traded for a more physically meaningful set: $m_h$, $m_H$, $m_A$, $m_{H^\pm}$, $m_{12}^2$, $\alpha$, and $\tan \beta$. Here, $ m_h $ and $ m_H $ denote the masses of two neutral scalar Higgs bosons, $ m_A $ represents the pseudoscalar Higgs mass, and $ m_{H^\pm} $ is the mass of the charged Higgs boson.  In addition, $ \alpha $ denotes the mixing angle of two neutral Higgs, while $ \tan \beta $ is defined as the ratio of the VEVs of $ \Phi_2 $ and $ \Phi_1 $, expressed as $\tan \beta  \equiv {v_2}/{v_1}$. 
The lighter $h$ and heavier $H$ are orthogonal combinations of $\phi_1$ and $\phi_2$ and can be given by:
\begin{equation}  
\begin{aligned}  
    h = \phi_1 \sin\alpha - \phi_2 \cos\alpha \,,\ H=\phi_1\cos\alpha+\phi_2\sin\alpha\,.
\end{aligned}  
\end{equation} 
The masses $m_{H^\pm}$ and $m_A$ are specified as follows \cite{Branco:2011iw}:
\begin{align}
m_{H^{\pm}}=&\sqrt{\frac{m_{12}^2-(\lambda_{4}+\lambda_{5})v_1v_2}{v_1v_2}(v_1^2 + v_2^2)}\,, \\ 
    m_A=&\sqrt{\frac{m_{12}^2-2\lambda_{5}v_1v_2}{v_1v_2}(v_1^2 + v_2^2)}\,.
\end{align}
It follows that, under the assumption $ m_{H^{\pm}} = m_A $, the parameters $ \lambda_4 $ and $ \lambda_5 $ are equal.

In the 2HDM-I, the Yukawa couplings of the up-type quarks $ u_R^i $, the down-type quarks $ d_R^i $, and the leptons $ e_R^i $ are all assumed to couple exclusively to $ \Phi_2 $, effectively eliminating all tree-level FCNCs \cite{Pich:2009sp, Tuzon:2010vt}.
Consequently, the Yukawa terms for the 2HDM-I Lagrangian based on the mass-eigenstate can be expressed as \cite{Aoki:2009ha}
\begin{widetext}
\begin{align}
\label{lg}
\mathcal{L}_{\mathrm{Yukawa}} =& \quad -\sum_{f=u,d,\ell}\frac{m_{f}}{v}\left(\xi_{h}^{f}\bar{f}fh+\xi_{H}^{f}\bar{f}fH-i\xi_{A}^{f}\bar{f}\gamma_{5}fA\right) \notag\\ 
& \quad -\left\{\frac{\sqrt{2}V_{ud}}{v} \bar{u}\left(m_{u}\xi_{A}^{u}\mathrm{P}_{L}+m_{d}\xi_{A}^{d}\mathrm{P}_{R}\right)dH^{+}+\frac{\sqrt{2}m_{\ell}\xi_{A}^{\ell}}{v} \bar{\nu}_{L}\ell_{R}H^{+}+\mathrm{h.c.}\right\}\,,
\end{align}
\end{widetext}
where $P_{R,L} = (1 \pm \gamma _5)/2$, $\xi _h^{u,d,\ell} =\cos\alpha/\sin\beta$, $\xi _H^{u,d,\ell} = \sin\alpha/\sin\beta$, $\xi _A^{u} = \mathrm{cot}\beta$ and $\xi _h^{d,\ell} = -\mathrm{cot}\beta$\,.
The couplings of Higgs to fermions and bosons in THDM-I are similar to those in SM, except for extra coefficients \cite{Carena:2002es}. 
 Combined with Eq.~\eqref{lg}, the couplings of the light and heavy neutral Higgs bosons ($ h $ and $ H $) to the vector bosons $ VV $ ($ VV = WW, ZZ $) in THDM-I, $C_{hVV}$ and $C_{HVV}$, can be derived as follows:
\begin{equation}
\begin{split}
C_{hVV} &= C_{hVV}^{\mathrm{SM}} \times \sin(\beta-\alpha)\,, \\ 
C_{HVV} &= C_{HVV}^{\mathrm{SM}} \times \cos(\beta-\alpha)\,. 
\end{split}
\label{eq9}
\end{equation}

The couplings of $ h $ and $ H $ to fermions in THDM-I, $C_{hff}$ and $C_{Hff}$, can be derived as follows:
\begin{equation}
\begin{split}     
C_{hff} &= C_{hff}^{\rm SM} \times \cos\alpha/\sin\beta\,, \\
C_{Hff} &= C_{Hff}^{\rm SM} \times \sin\alpha/\sin\beta\,.
\end{split}
\label{tth}
\end{equation}
The decay width of the light Higgs to $\gamma\gamma$, denoted as $\Gamma(h\to\gamma\gamma)$, is given as \cite{Djouadi:2005gj, Posch:2010hx}
%\begin{widetext}
\begin{align}
&\Gamma(h\to\gamma\gamma)=\frac{G_F\alpha^2m_h^3}{128\sqrt{2}\pi^3} \times \nonumber\\
&\bigg|\sum_{f\in\{t,b\}}C_{f}Q_f^2N_cA_{1/2}(\tau_f)+C_{W}A_1(\tau_W)+C_{H^{\pm}}A_0(\tau_{H^\pm})\bigg|^2.
\label{eq10}
\end{align}
%\end{widetext}
Here, $C_W = C_{hWW}/C_{hWW}^{\rm SM}$ represents the ratio of Higgs to $ W $ boson coupling in the 2HDM relative to that of the SM, and $C_f = C_{hff}/C_{hff}^{\rm SM}$ signifies the corresponding ratio for fermion couplings. Furthermore, $C_{H^{\pm}} = -m_W/gm^2_{H^{\pm}} \times C_{hH^+H^-}$, where $ C_{hH^+H^-} $ is the coupling between $ h $ and $ H^\pm $, the detailed information can be obtained from Ref.~\cite{Bai:2012ex}. With $ \tau_i = {m_h^2}/{(4m_i^2)} $ (for $ i = f, W, H^\pm $) denotes the normalized mass terms, the functions $ A_0, A_{1/2}, A_1 $ are defined as follows:
\begin{equation}
\begin{aligned}
A_{0}(\tau_{H^{\pm}})& =-[\tau_{H^{\pm}}-f(\tau_{H^{\pm}})]\tau_{H^{\pm}}^{-2} , \\
A_{1/2}(\tau_f)& =2[\tau_f+(\tau_f-1)f(\tau_f)]\tau_f^{-2} , \\
A_{1}(\tau_W)& =-[2\tau_W^{2}+3\tau_W+3(2\tau_W-1)f(\tau_W)]\tau_W^{-2} , 
\end{aligned}
\end{equation}
where
\begin{align}
f(\tau)=\left\{\begin{array}{cc}\arcsin^2\sqrt{\tau}&,\quad\tau\leq1 \\ 
 -\frac{1}{4}\left[\log\frac{1+\sqrt{1-\tau^{-1}}}{1-\sqrt{1-\tau^{-1}}}-i\pi\right]^2&,\quad\tau>1\end{array}.\right.
\end{align}

\section{\label{sec:scan}Parameters space scan and discussion}
\subsection{\label{scan}Parameters space scan and constraints}
We use the Mathematica package \textsf{SARAH-4.15.2} \cite{Staub:2009bi, Staub:2010jh, Staub:2012pb, Staub:2013tta} to generate the model file and then use the package \textsf{SPheno-4.0.5} \cite{Porod:2003um, Porod:2011nf} to calculate the particle spectrum. The seven input parameters are $m_h$, $m_H$, $m_A$, $m_{H^{\pm}}$, $m_{12}^2$, $\alpha$, and $\tan\beta$. Since one of the CP-even Higgs bosons should be the SM-like, we denote $H$ as the SM-like Higgs with $m_H = 125\ \mathrm{GeV}$. The mass of $ h $ is set to 95 GeV. The other parameters are explored within the following ranges:
\begin{equation}  
\begin{aligned}  
    \alpha &\in [-\pi/2,\ \pi/2], \quad \\ \tan\beta &\in [0,\ 20],  \quad \\
    m_{12}&\in [-10 \text{ TeV},\ 10 \text{ TeV}] , \\
    m_{H^{\pm}}= m_{A} &\in [150 \text{ GeV},\ 1000 \text{ GeV}].
\end{aligned}  
\end{equation}

%There are numerous constraints on 2HDM-I. 
For the experimental constraints, we consider the following:
\begin{itemize}

    \item Constraints from B physics, $B_d \to \mu^+\mu^-$, $B_s \to \mu^+\mu^-$, and $B \to X_s\gamma$ \cite{ATLAS:2018cur, CMS:2022mgd, LHCb:2021vsc, HFLAV:2022esi, ParticleDataGroup:2022pth} \add{at the $2\sigma$ level}:

    \begin{equation}
    \begin{aligned}
    \mathrm{BR}&(B_{d}\to\mu^{+}\mu^{-})<2.1\times10^{-10} ,\\
    \mathrm{BR}&(B_{s}\to\mu^{+}\mu^{-})=(\add{3.34\pm0.58})\times10^{-9} , \\
    \mathrm{BR}&(B\to X_{s}\gamma)=(3.32\pm0.3)\times10^{-4}.
    \end{aligned}
    \end{equation}
    The package \textsf{SuperIso-v4.1} \cite{Mahmoudi:2007vz, Mahmoudi:2008tp} is used to calculate the branching ratios related to B physics.

    \item Constraints from direct Higgs searches in the diphoton channel, such as $e^+e^- \to ZH\ (\to \gamma\gamma)$ at LEP \cite{ALEPHDELPHIL3andOPALCollaborations:2002sal} and $pp \to h \to \gamma\gamma$ at the LHC \cite{ATLAS:2014jdv, CMS:2015ocq, CMS:2018cyk, CMS:2024yhz}, are taken into account.
    Additional constraints from other search channels are evaluated at the 95\% confidence level using the \textsf{HiggsBounds-5.10.0} package \cite{Bechtle:2020pkv}.

    \item Constraints from 125 GeV SM-like Higgs data. We use the \textsf{HiggsSignal-2.6.0} package \cite{Bechtle:2020uwn} to calculate a $\chi^2$ to evaluate the compatibility of the 125 GeV Higgs with the Standard Model Higgs. For 107 degrees of freedom, a $\chi^2$ value less than 124.3 is required at the 95\% confidence level.
    value using the peak-centered $\chi^2$ method

    \item Constraints from electroweak precision observables \cite{Peskin:1990zt, Peskin:1991sw} requiring that the oblique parameters $ S $, $ T $, and $ U $ to  matching the global fit results as detailed in \cite{ParticleDataGroup:2022pth}:
    \begin{equation}
    \begin{aligned}
    S&=-0.04\pm0.10\,,\\  T&=0.01\pm0.12\,, \\ U&=-0.01\pm0.09\, ,
    \end{aligned} 
    \end{equation}
    
    where the correlation coefficients are given as $\rho_{ST} = 0.92$, $\rho_{SU} = -0.80$, and $\rho_{TU} = -0.93$. %with the requirement that $\chi_{ST, SU, TU} < 5.99$.

\end{itemize}

For theoretical constraints, we account for the following:
\begin{itemize}
    
    \item Constraints from unitarity on scalar parameters requiring that the scalar couplings $\lambda_i$ (where $i=1, 2, \ldots, 5$) to satisfy specific conditions \cite{Kanemura:1993hm, Akeroyd:2000wc, Arhrib:2000is}:
    \begin{widetext}
    \begin{equation}
    \begin{aligned}
        &\left| 2\lambda_3 + \frac{\lambda_5}{2} \right| \leq 8\pi , \quad
        \left| 2\lambda_3 \pm \lambda_4 - \frac{\lambda_5}{2} \right| \leq 8\pi , \quad
        \left| 2\lambda_3 + \lambda_4 - \frac{\lambda_5}{2} \right| \leq 8\pi , \\
        &\left| 2\lambda_3 - \lambda_4 - \frac{5\lambda_5}{2} \right| \leq 8\pi ,  \quad
        \left| 2\lambda_3 + 2\lambda_4 - \frac{\lambda_5}{2} \right| \leq 8\pi ,\\
        &\left| \lambda_1 + \lambda_2 + 2\lambda_3 \pm \sqrt{(\lambda_1 - \lambda_2)^2 + \frac{1}{4}\lambda_5^2} \right| \leq 8\pi ,\\
        &\left| \lambda_1 + \lambda_2 + 2\lambda_3 \pm \sqrt{(\lambda_1 - \lambda_2)^2 + \frac{1}{4}(-2\lambda_4 + \lambda_5)^2} \right| \leq 8\pi ,\\
        &\left| 3(\lambda_1 + \lambda_2 + 2\lambda_3) \pm \sqrt{9(\lambda_1 - \lambda_2)^2 + \left(4\lambda_3 + \lambda_4 + \frac{1}{2}\lambda_5\right)^2} \right| \leq 8\pi .    
    \end{aligned}
    \end{equation}
    \end{widetext}

    \item Constraints from vacuum stability, as outlined in \cite{ElKaffas:2006gdt}:
    \begin{equation}
    \begin{aligned}
    \lambda_{1},\lambda_{2} > 0\,,\ \lambda_{3}>-\sqrt{\lambda_{1}\lambda_{2}}  \, , \\ \lambda_3+\min[0,\lambda_4-|\lambda_5|]>-\sqrt{\lambda_1\lambda_2}\, .
    \end{aligned}
    \end{equation}

    \item Constraints from perturbativity, as outlined in \cite{Branco:2011iw}, requiring that the absolute values of the scalar couplings $\lambda_i$ (for $i = 1, 2, \ldots, 5$) to be less than $8\pi$, specifically, $|\lambda_i| < 8\pi$.

\end{itemize}
\subsection{\label{scan}Surviving condition and discussion}

\begin{figure*}[!tbp] 
\centering 
\includegraphics[width=0.8\linewidth]{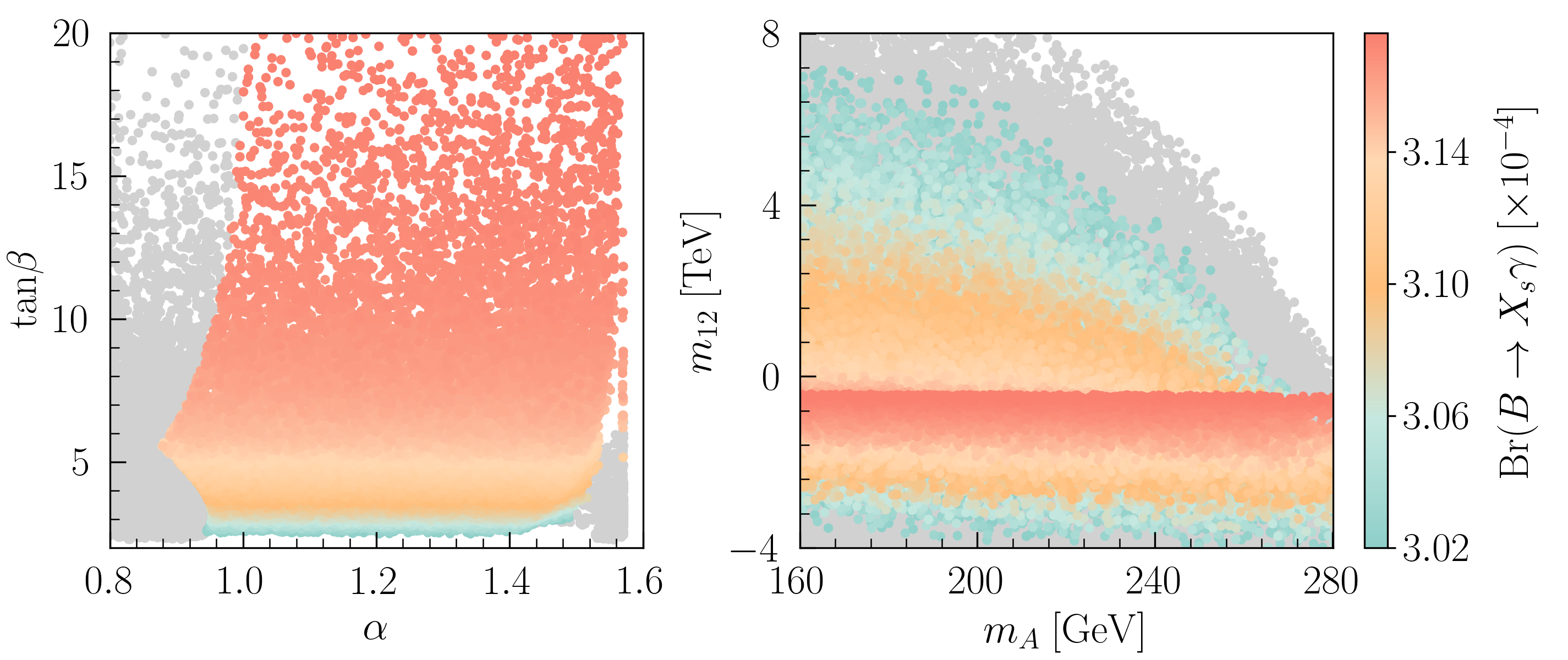}
\caption{\label{f1} 
Surviving samples in the $\tan\beta$ versus $\alpha$ (left) and $m_{12}$  versus $m_A$ planes (right) with colors indicating the branching ratio of $B\to X_s \gamma$ and the gray samples are excluded by Higgs direct search constraints.
}
\end{figure*}

In Fig.~\ref{f1}, the surviving samples are shown in the $\tan\beta$ versus $\alpha$ (left) and  $m_{12}$  versus $m_A$ planes (right), with colors indicating the branching ratio of $B\to X_s \gamma$ and the gray samples are excluded by Higgs direct search constraints.
The results of the parameter space scan indicate that direct searches for BSM Higgs bosons impose strong constraints on $\alpha$, excluding all samples with $\alpha \lesssim 0.9$.
In addition, one can observe that the branching ratio $\text{Br}(B \to X_s \gamma)$ strongly depends on $\tan\beta$. The surviving samples are required to satisfy $\tan\beta \gtrsim 2.6$ to be consistent with experimental constraints. Moreover, smaller values of $m_A$ and $|m_{12}|$ tend to enhance $\text{Br}(B \to X_s \gamma)$ at fixed $\tan\beta$. 
To achieve a larger $\text{Br}(B \to X_s \gamma)$ generally requires a larger $\tan\beta$ and a smaller $|m_{12}|$.

In Fig.~\ref{f2}, the surviving samples are shown in the $\tan\beta$ versus $\alpha$ (upper three, lower left, and lower middle) and $\mu_{\gamma\gamma}$ versus $\alpha$ (lower right) planes. The color coding indicates the following observables: the cross section $\sigma(gg \to h)$ (upper left), the branching ratio $\text{Br}(h \to \gamma\gamma)$ (upper middle), the cross section $\sigma(gg \to h \to \gamma\gamma)$ (upper right and lower right, with CMS and ATLAS measurements marked), the squared reduced top-Higgs coupling $(C_{tth}/C_{tth}^{\rm SM})^2$ normalized to its SM value in the 2HDM-I (lower left), and the diphoton signal strength $\mu_{\gamma\gamma}$ (lower middle).
The measured signal strength and associated uncertainties for the 95 GeV diphoton excess observed by CMS and ATLAS are also indicated in the lower right panel of Fig.~\ref{f2}.
The signal strength $\mu_{\gamma\gamma}$ is a well-known observable for probing BSM phenomenology in the diphoton channel. It can be parameterized as
\begin{align}
    \mu_{\gamma\gamma}(h) = \frac{\sigma_{\mathrm{BSM}}(gg\to h) \times \mathrm{Br_{BSM}}(h\to \gamma\gamma)}{\sigma_{\mathrm{SM}}(gg\to h_{95}) \times \mathrm{Br_{SM}}(h_{95}\to \gamma\gamma)},
\end{align}
where $h$ ($h_{95}$) is the 95 GeV Higgs in BSM (SM), $\sigma_{\mathrm{BSM}}(gg\to h)$ ($\sigma_{\mathrm{SM}}(gg\to h_{95})$) denotes the production cross section of the light Higgs in the BSM (SM), and $\mathrm{Br}_{\mathrm{BSM}}(h \to \gamma\gamma)$ ($\mathrm{Br}_{\mathrm{SM}}(h_{95} \to \gamma\gamma)$) is the corresponding diphoton branching ratio.
In the SM, the cross section for $gg \to h_{95}$ at the 14 TeV LHC is 79.83 pb~\cite{LHCHiggsCrossSectionWorkingGroup:2011wcg}, and the branching ratio for $h_{95} \to \gamma\gamma$ is $1.39\times 10^{-3}$~\cite{LHCHiggsCrossSectionWorkingGroup:2013rie}.
The measured diphoton signal strengths are $\mu_{\gamma\gamma}^{\mathrm{ATLAS}} = 0.18^{+0.1}_{-0.1}$ from ATLAS \cite{ATLAS:2023jzc} and $\mu_{\gamma\gamma}^{\mathrm{CMS}} = 0.33^{+0.19}_{-0.12}$ from CMS \cite{CMS:2024yhz}. 
Several conclusions can be drawn from Fig.~\ref{f2}:
\begin{itemize}

    \item From the upper left and lower left panels of Fig.~\ref{f2}, one can observe that both the cross section $\sigma(gg \to h)$ and the reduced top-Higgs coupling $(C_{tth}/C_{tth}^{\rm SM})^2$ exhibit similar trends with respect to $\alpha$ and $\tan\beta$. They are approximately proportional to $\cos^2\alpha$ and inversely proportional to $\sin^2\beta$.  
    This behavior of the reduced top-Higgs coupling is consistent with Eq.~\eqref{tth}.  
    In the 2HDM-I, the cross section $\sigma(gg \to h)$ can be obtained by rescaling the SM prediction by a factor of $\cos^2\alpha/\sin^2\beta$, as the loop contribution is dominated by the top quark, which is typically the case.

    \item From the upper middle panel of Fig.~\ref{f2}, it is evident that the branching ratio of the lighter Higgs boson decaying into diphoton, $\text{Br}(h \to \gamma\gamma)$, primarily depends on the parameters $\alpha$ and $\tan\beta$. 
    The branching ratio reaches its minimum value of about $10^{-6}$ when $\sin(\beta-\alpha)\approx0$.  
    This behavior can be understood from the partial decay width $\Gamma(h \to \gamma\gamma)$, which is described by Eq.~\eqref{eq10}, where the loop functions take typical values $A_0(\tau_{H^\pm}) \approx 0.34$, $A_{1/2}(\tau_f) \approx 1.4$, and $A_1(\tau_W) \approx -7.6$ in our parameter scan. 
    The dominant contribution to $\Gamma(h \to \gamma\gamma)$ arises from the $W$ boson loop. When $\sin(\beta-\alpha)$ approaches zero, the coupling of $h$ to the $W$ boson becomes suppressed (see Eq.~\eqref{eq9}), leading to a negligible $W$ contribution to the diphoton decay width.
    One can also note that $\text{Br}(h \to \gamma\gamma)$ approaches 1 when $\cos\alpha\approx0$. In this region, the decay of $h$ to fermions is highly suppressed, consistent with Eq.~\eqref{tth} which shows that the $h$-fermion coupling approaches 0.

    \item From the upper right and lower middle panels of Fig.~\ref{f2}, it can be seen that the cross section $\sigma(gg \to h \to \gamma\gamma)$ reaches up to 0.065 pb around $\alpha\approx1.5$ and $\tanb\approx5$, corresponding to a diphoton signal strength of $\mu_{\gamma\gamma} \approx 0.59$.  
    The 95 GeV diphoton excess observed by CMS and ATLAS can be accommodated within the $1\sigma$ range for a wide range with $0.95\lesssim\alpha\lesssim1.55$, as shown in the lower middle and right panels of Fig.~\ref{f2}.

\begin{figure*}[!tbp] 
\centering 
\includegraphics[width=\linewidth]{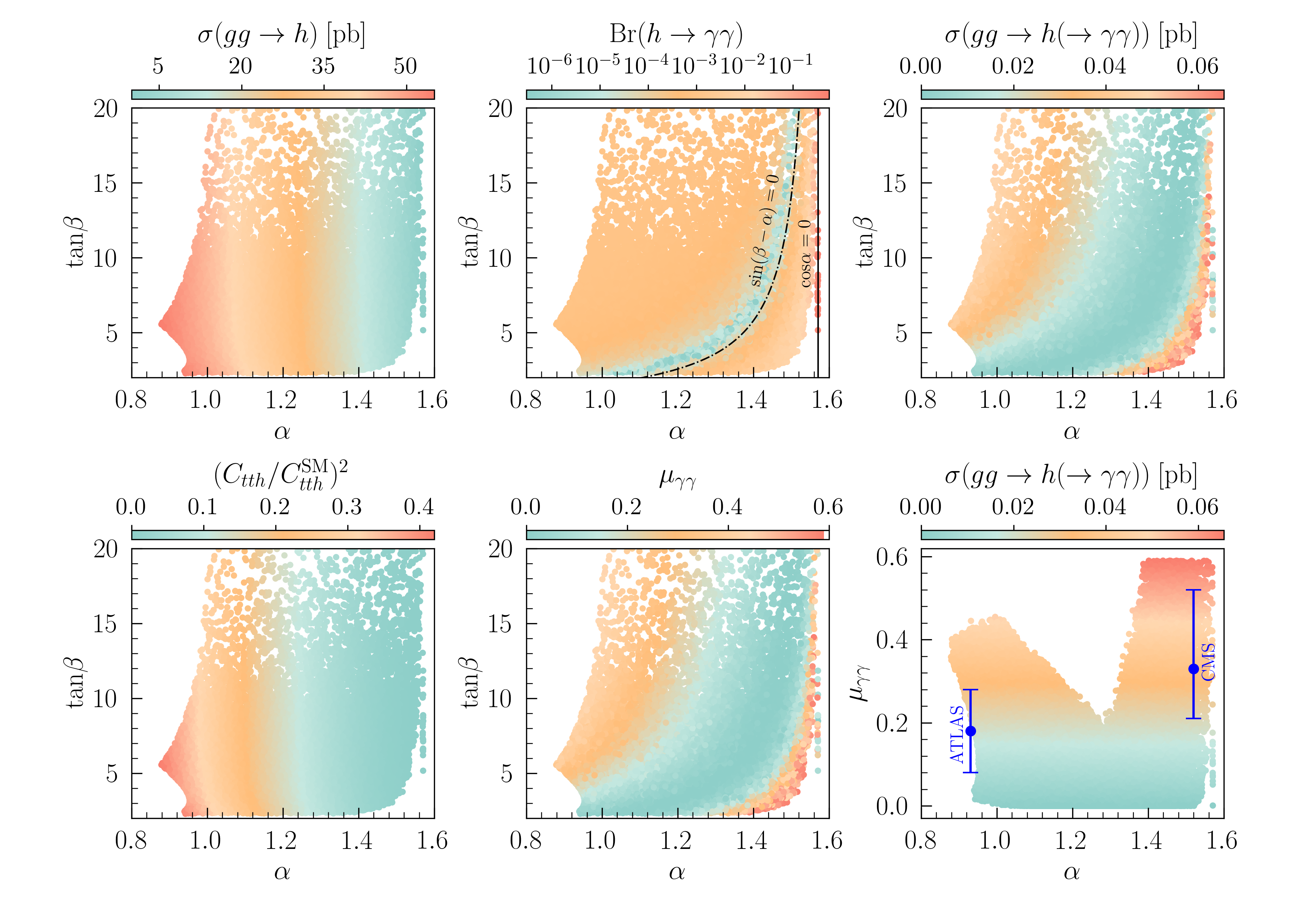}
\vspace{-30pt}
\caption{\label{f2} 
Surviving samples in the $\tan\beta$ versus $\alpha$ (upper three, lower left, and lower middle) and $\mu_{\gamma\gamma}$ versus $\alpha$ (lower right) planes. Colors indicate the following observables: the cross section $\sigma(gg \to h)$ (upper left), the branching ratio $\text{Br}(h \to \gamma\gamma)$ (upper middle), the cross section $\sigma(gg \to h \to \gamma\gamma)$ (upper right and lower right), the reduced top-Higgs coupling $(C_{tth}/C_{tth}^{\rm SM})^2$ normalized to the SM value in the 2HDM-I (lower left), and the diphoton signal strength $\mu_{\gamma\gamma}$ (lower middle). In the lower right panel, markers represent the measured signal strengths and uncertainties for the 95 GeV diphoton excess reported by CMS \cite{CMS:2024yhz} and ATLAS \cite{ATLAS:2023jzc}.
}
\end{figure*}

\end{itemize}

\subsection{\label{scan}The discussion on muon g-2}

Recently, the Fermi National Accelerator Laboratory (FNAL) released its latest measurement of the muon magnetic moment, which is more precise and remains consistent with its previous measurements as well as those from the Brookhaven National Laboratory (BNL)~\cite{Muong-2:2006rrc, Muong-2:2021ojo, Muong-2:2023cdq, Muong-2:2025xyk}. The latest experimental result for the muon magnetic moment ($a_{\mu}^{\rm exp}$) is
\begin{align}
    a_{\mu}^{\rm exp} = 1165920715(145)\times10^{-12}.
\end{align}

On the theory side, the Standard Model (SM) prediction for the muon magnetic moment has also been recently updated. Due to differences in calculation approaches, particularly in the evaluation of the leading-order (LO) hadronic vacuum polarization (HVP) contribution, the predicted value has shifted significantly.
In the White Paper 2020 (WP20), the LO HVP contribution was estimated using a data-driven dispersive method, resulting in~\cite{Aoyama:2020ynm}
\begin{align}
    a_{\mu}^{\rm SM-WP20} = 116591810(43)\times10^{-11},
\end{align}
which leads to a discrepancy with the experimental measurement exceeding $5\sigma$.
In contrast, the White Paper 2025 (WP25) adopted a lattice-QCD-based approach to evaluate the LO HVP contribution. The updated SM prediction reads~\cite{Aliberti:2025beg}
\begin{align}
    a_{\mu}^{\rm SM-WP25} = 116592033(62)\times10^{-11},
\end{align}
which is in agreement with the experimental result within $1\sigma$.

\begin{figure*}[!tbp] 
\centering 
\includegraphics[width=0.7\linewidth]{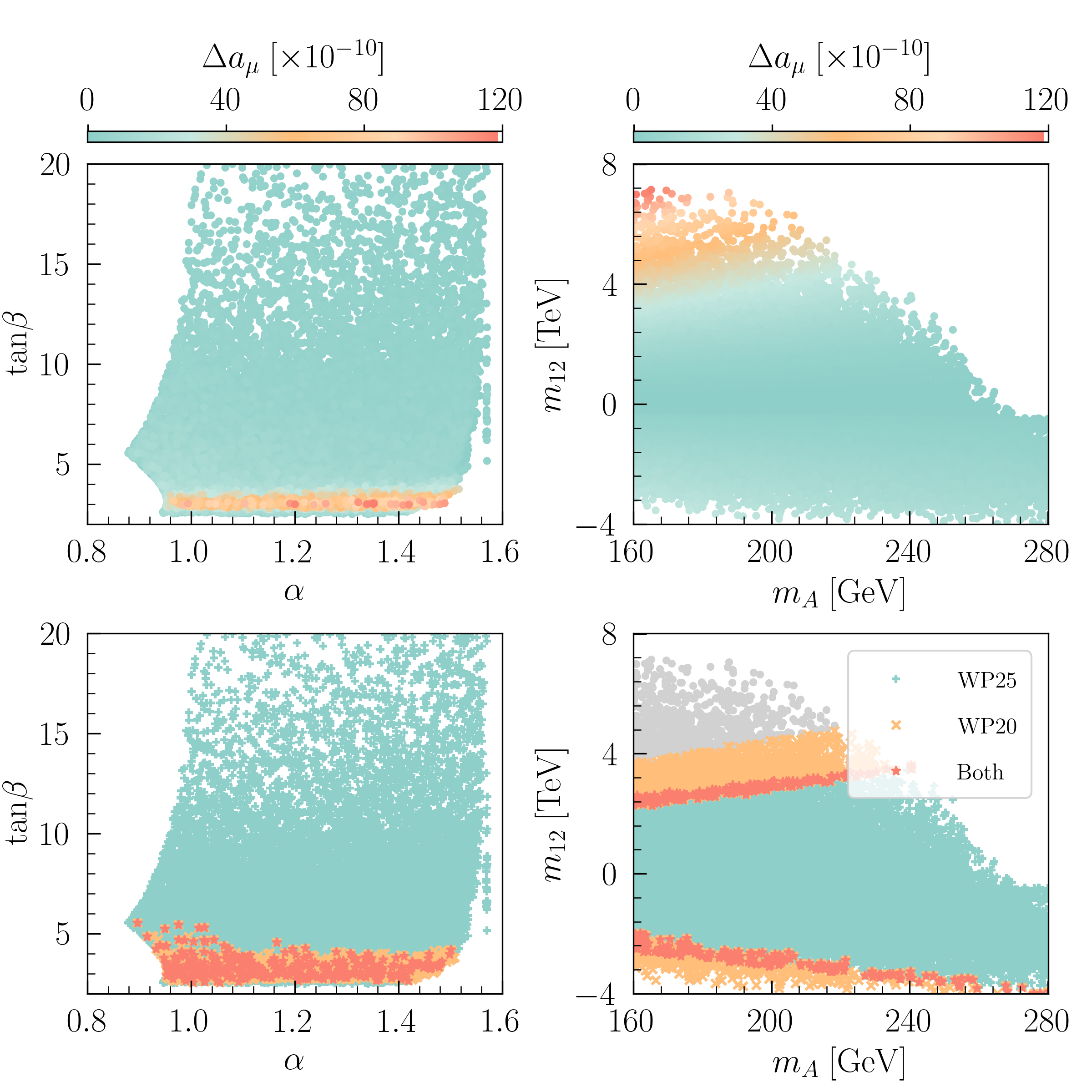}
\caption{\label{f3}
Surviving samples in the $\tan\beta$ versus $\alpha$ (left) and $m_{12}$ versus $m_A$ (right) planes. The color gradient in the upper panels represents the muon anomalous magnetic moment $\Delta a_{\mu}$, evaluated at the two-loop level. In the lower panels, blue plus signs (+), orange crosses ($\times$), and red stars (*) denote samples consistent with the WP25 results, the WP20 results, and both, respectively, within the $2\sigma$ range. Gray points correspond to samples that are incompatible with either result.
}
\end{figure*}

Fig.~\ref{f3} displays the surviving samples in the $\tan\beta$ versus $\alpha$ (left panel) and $m_{12}$ versus $m_A$ (right panel) planes. The color map in the upper panels represents the anomalous muon magnetic moment $\Delta a_{\mu}$, computed at the two-loop level using the \textsf{GM2Calc-2.2.0} package~\cite{Athron:2015rva, Athron:2021evk}. In the lower panels, blue crosses (+), orange crosses ($\times$), and red stars (*) denote the samples that satisfy the WP25 result, the WP20 result, and both, respectively, at the $2\sigma$ level. Gray points do not satisfy either constraint.

From the upper two panels of Fig.~\ref{f3}, it is evident that a larger $|m_{12}|$ tends to predict a larger $\Delta a_{\mu}$, whereas a heavier $m_A$ leads to a smaller $\Delta a_{\mu}$. In general, samples satisfying $|m_{12}| \lesssim 13\, m_A + 500$ GeV can yield a $\Delta a_{\mu}$ value consistent with experimental observations. 
In addition, $\Delta a_{\mu}$ is also influenced by $\tanb$ and has its maximum value when $\tanb\approx3$.
This effect arises from additional corrections of the pseudoscalar $A$ induced by one-loop diagrams \cite{Zhou:2001ew, Assamagan:2002kf, Davidson:2010xv, Han:2022juu}, which are expressed as
\begin{align}
    \Delta a_\mu^{1 \rm L}&=\frac{m_\mu m_\tau\rho^2}{8\pi^2} \left[\frac{(\log\frac{m_H^2}{m_\tau^2}-\frac{3}{2})}{m_H^2}-\frac{\log(\frac{m_A^2}{m_\tau^2}-\frac{3}{2})}{m_A^2}\right] ,
\end{align}
where $\rho$ is the coupling parameter for $\mu$ and $\tau$ to $H$ and $A$. It can be seen that $\Delta a_\mu$ gets a positive contribution at the one-loop level only if $m_A$ is greater than $m_H$. In our calculations, $m_A$ is not significantly larger than $m_H$, so the term inside the square brackets is about $10^{-4}$, and the term before the square brackets is about $10^{-8}$. Consequently, the net contribution from one-loop processes is about $10^{-12}$. The two-loop contributions to $\Delta{a_{\mu}}$ are divided into bosonic ($a_{\mu}^B$) and fermionic ($a_{\mu}^F$) parts, as detailed in \cite{Cherchiglia:2016eui, Cherchiglia:2017uwv, Li:2020dbg}, which given by
\begin{equation}
    \Delta{a_{\mu}^{2 \rm L}} = a_{\mu}^{ \rm B} + a_{\mu}^{ \rm F}\,.
\end{equation}
The pseudoscalar $A$ and charged Higgs boson $H^{\pm}$ also contribute to$a_{\mu}^{\rm F}$ and $a_{\mu}^{\rm B}$ via loop effects, with contributions decreasing as their masses increase \cite{Cherchiglia:2016eui}.
The dominant contribution to $\Delta{a_{\mu}}$ comes from $a_{\mu}^{\rm B}$, which consists of three parts and have
\begin{equation}
    a_{\mu}^{ \rm B}=a_{\mu}^{ \rm EW}+a_{\mu}^{ \rm Yuk}+a_{\mu}^{ \rm non-Yuk}.
\end{equation}
The component $a_{\mu}^{\rm EW}$ corresponds to contributions from the SM-like Higgs, $a_{\mu}^{\rm non-Yuk}$ represents QCD contributions without new Yukawa couplings and $a_{\mu}^{\rm Yuk}$ accounts for QCD contributions with new Yukawa couplings. The latter has the form
\begin{widetext}
\begin{align}
    a_{\mu}^{ \rm Yuk}= \frac{\alpha_{em}^2m_{\mu}^2}{574\pi ^2 c_W^4s_W^4m_Z^4} \Big[ a_{0, 0}^0 + a_{5, 0}^0 \Lambda_5 + \big( a_{0, 0}^1\frac{\mathrm{tan}^2\beta-1}{\mathrm{tan}\beta} + a_{5, 0}^1\frac{\mathrm{tan}^2\beta-1}{\mathrm{tan}\beta} \Lambda_5 \big)\mathrm{cos}(\beta-\alpha)\Big],
\end{align}
\end{widetext}
where $\Lambda_5$ is defined as $2m_{12}^2/v^2 \sin \beta \cos \beta$, and the coefficients $a_{0, 0}^0$, $a_{0, 0}^1$, $a_{5, 0}^0$, and $a_{5, 0}^1$ are given in the Appendix of Ref.~\cite{Cherchiglia:2016eui}. In our analysis, $m_{12}^2$ is significantly larger than both $\tan\beta$ and $v^2$, causing $a_{\mu}^{\rm Yuk}$ to be nearly proportional to $m_{12}^2$. Consequently, $\Delta a_\mu$ increases with increasing $m_{12}^2$.

From the lower panels of Fig.~\ref{f3}, one can observe that the WP25 result can be satisfied across the entire $\tan\beta$ range, whereas the WP20 result prefers smaller values, typically $\tan\beta \lesssim 5$. In the $m_{12}$ versus $m_A$ plane, most samples with $-3\,\mathrm{TeV} \lesssim m_{12} \lesssim 3\,\mathrm{TeV}$ are consistent with the WP25 result. Additionally, a few rare samples in the region $18\,m_A - 400\,\mathrm{GeV} \lesssim |m_{12}| \lesssim 18\,m_A + 600\,\mathrm{GeV}$ also satisfy the WP25 constraint.

\section{\label{sec:result}Future collider simulations and discussion}
\subsection{\label{MC}Monte Carlo (MC) simulations}
\begin{figure*}[!tbp] 
\centering 
\includegraphics[width=0.9\linewidth]{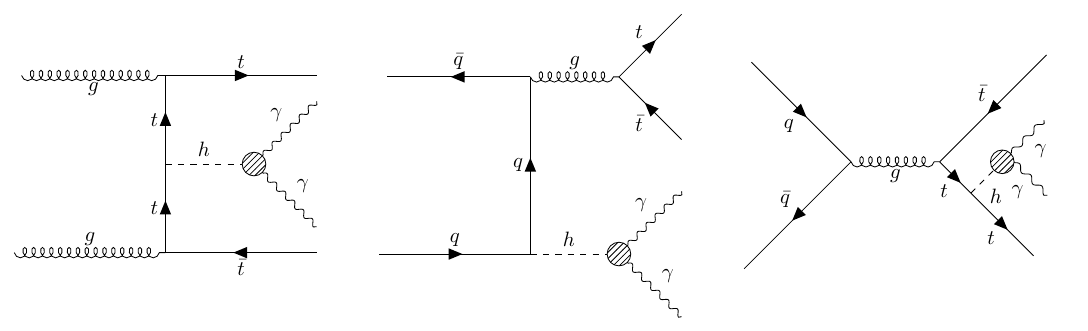}
\caption{\label{feyn} 
Main Feynman Diagrams for the Process $pp \to t\bar{t}h(\to \gamma\gamma)$ at LO in 2HDM-I.
}
\end{figure*}

In this study, we focus on the process $pp \to t(\to W^+ b)\, \bar{t}(\to W^- \bar{b})\, h(\to \gamma\gamma)$, where the Higgs boson $h$ is produced in association with a top-quark pair and subsequently decays into two photons. Each top quark decays into a $W$ boson and a $b$ quark.
The LO Feynman diagrams for the process $pp \to t\bar{t}h(\to \gamma\gamma)$ in the 2HDM-I are shown in Fig.~\ref{feyn}. The decay of $h$ into a diphoton proceeds only via loop processes, which significantly suppresses the corresponding cross section.
Nevertheless, the diphoton final state offers a relatively clean experimental signature at hadron colliders. In particular, the invariant mass of the diphoton system gives rise to a narrow resonance peak around the mass of $h$, making the signal easier to isolate.
Moreover, the presence of the associated top-quark pair further suppresses SM backgrounds, as the top quark decays into a $W$ boson and a $b$ quark, leaving additional detectable jets in the final state.
In addition, the $W$ boson can decay leptonically, producing a charged lepton and a neutrino; the latter escapes detection and contributes to missing transverse momentum, while the charged lepton provides a useful trigger for event selection.

\begin{figure*}[!tbp] 
\centering 
\includegraphics[width=\linewidth]{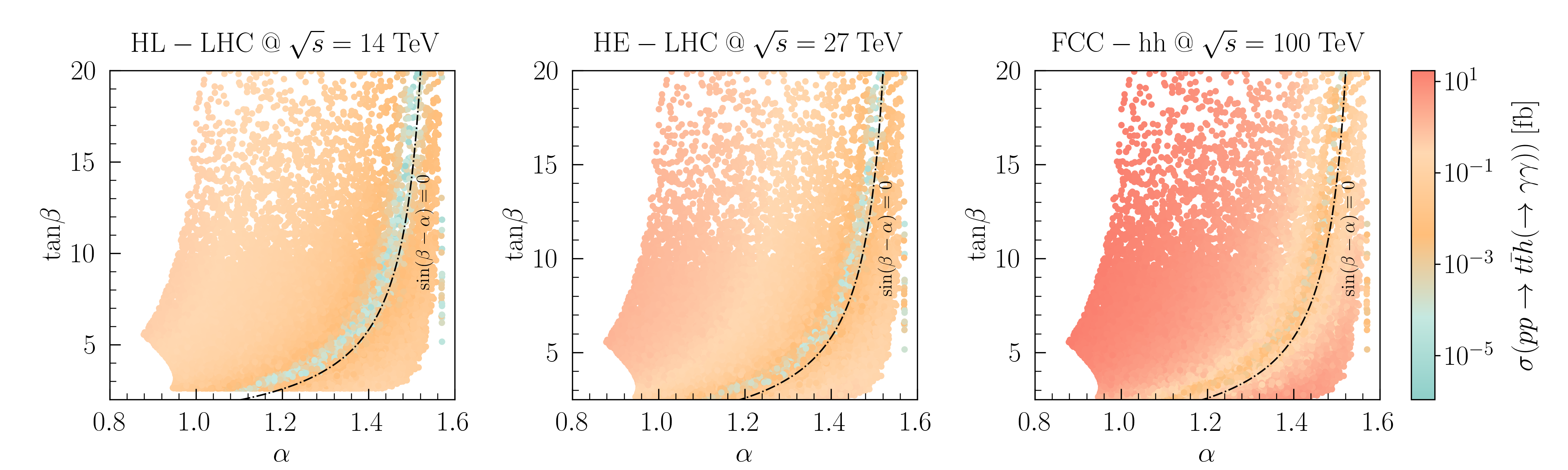}
\vspace{-10pt}
\caption{\label{f5} 
Surviving samples on the $\tan\beta$ versus $\alpha$ plane, with colors indicating the cross section $\sigma(pp\to t\bar{t}h(\to \gamma\gamma))$ at 14 TeV High-Luminosity LHC (HL-LHC) (left), 27 TeV High-Energy LHC (HE-LHC) (middle), and 100 TeV Future Circular Collider (FCC-hh) (right). 
}
\end{figure*}

In Fig.~\ref{f5}, the surviving samples are presented in the $\tan\beta$ versus $\alpha$ plane, with color indicating the cross section $\sigma(pp \to t\bar{t}h \to \gamma\gamma)$ at the 14 TeV High-Luminosity LHC (HL-LHC) (left), the 27 TeV High-Energy LHC (HE-LHC) (middle), and the 100 TeV Future Circular Collider (FCC-hh) (right).
The cross section $\sigma(pp \to t\bar{t}h \to \gamma\gamma)$ at the HL-LHC is calculated as
\begin{align}
\sigma(pp\to t\bar{t}(h\to\gamma\gamma))=&\ |\sigma(pp\to t\bar{t}H)|_{m_{h}=95 \mathrm{GeV}}^{\rm SM}\nonumber \\ &\times (C_{t\bar{t}h}/C_{t\bar{t}h}^{\rm SM})^{2}\times Br(h\to\gamma\gamma),
\end{align}
where $|\sigma(pp\to t\bar{t}H)|_{m_{h}=95 \mathrm{GeV}}^{\rm SM}$ is the SM cross section for the process $ pp \to t\bar{t}H $ with a Higgs mass of 95 GeV. The value at next-to-leading order (NLO) is reported to be 1268 fb, as given in  Ref.~\cite{LHCHiggsCrossSectionWorkingGroup:2011wcg}.
The cross sections at the HE-LHC and FCC-hh are calculated using the following rescaling formula:
\begin{align}
\sigma _{\rm HE-LHC(FCC-hh)} =&\  \sigma _{\rm HE-LHC(FCC-hh)}^{\rm P1} \nonumber\\ &\times \frac{C_{t\bar{t}h}^2 \times Br(h\to\gamma\gamma)}{(C_{t\bar{t}h}^{\rm P1})^2 \times Br^{\rm P1}(h\to\gamma\gamma)},
\end{align}
where $\sigma _{\rm HE-LHC}^{\rm P1}$ and $\sigma _{\rm FCC-hh}^{\rm P1}$ are cross sections for the Benchmark Point P1 at the 27 TeV HE-LHC and 100 TeV FCC-hh, calculated using \textsf{MadGraph5\_aMC@NLO\_v3.4.2} \cite{Alwall:2011uj, Alwall:2014hca}, and they correspond to 0.64 fb and 7.75 fb, respectively. Additionally, $C_{t\bar{t}h}^{\rm P1}$ and $Br^{\rm P1}(h\to\gamma\gamma)$ denote the reduced coupling and the branching ratio for this benchmark point. 
For the surviving samples, the maximum cross sections of the process $pp \to t\bar{t}(h \to \gamma\gamma)$ can reach 0.44 fb at the HL-LHC, 1.42 fb at the HE-LHC, and 17.21 fb at the FCC-hh.
The minimum value of the cross section occurs near $\mathrm{sin}(\beta-\alpha)\approx0$ because Br($h\to \gamma\gamma$) is particularly small there.
Detailed information on five representative benchmark points selected from the surviving samples is provided in Table~\ref{t00}.

\begin{table*}[!tbp] 
\caption{The detailed information of five benchmark points for surviving samples.}
\centering
    \setlength{\tabcolsep}{23pt}
    \resizebox*{1\textwidth}{!}{
    \begin{tabular}{@{\hspace{20pt}}cccccc@{\hspace{20pt}}}
    \toprule
            ~ & P1 & P2 & P3 & P4 & P5 \\ \midrule
    $\alpha$ & 1.08 & 0.95 & 1.34 & 0.96& 1.01  \\  
    tan$\beta$ & 7.19 & 4.28 & 2.90 & 3.61 & 4.33  \\  
    $m_{H^\pm}(m_A)$ [GeV]& 178 & 210 & 211 & 223 & 184  \\  
    $m_{12}$ [GeV]& -955 & -2099 & -1809 & 990 & 347  \\  
    $\lambda_{1}$ & 1.58 & 0.36 & 0.32 & 1.71 & 2.02  \\  
    $\lambda_{2}$ & 0.12 & 0.11 & 0.14 & 0.12 & 0.12  \\  
    $\lambda_{3}$ & 1.27 & 1.53 & 1.44 & 1.91 & 1.37 \\  
    $\lambda_{4}$ & -0.41 & -0.57 & -0.64 & -0.89 & -0.59 \\  
    $\lambda_{5}$ & -0.41 & -0.57 & -0.64 & -0.89 & -0.59  \\  
    T $[\times 10^{-2}]$& -1.5 & -1.5 & -1.7 & -1.5 & -1.5  \\  
    S $[\times 10^{-3}]$& 9.2 & 6.7 & 0.9 & 6.6 & 8.9 \\  
    U $[\times 10^{-4}]$& -4.4 & -5.1 & -6.4 & -5.7 & -4.7  \\  
    Br($h\to \gamma\gamma$) $[\times10^{-3}]$ &         0.69 & 0.44 & 1.09 & 0.33 &0.45  \\ 
    $\sigma(gg\to h)$ [pb]                              & 40.6 & 50.8 & 19.8 & 50.5 & 46.3 \\ 
    $\sigma(gg\to h(\to \gamma\gamma))$ [fb]            & 28.0 & 22.2 & 21.6 & 17.0 & 20.8 \\ 
    $\mu_{\gamma\gamma}$                                & 0.25 & 0.20 & 0.19 & 0.15 & 0.19 \\
    $(C_{htt}/C_{htt}^{\mathrm{SM}})^2$                 & 0.23 & 0.36 & 0.06 & 0.36 & 0.30 \\  
    $\sigma(pp\to t\bar{t}h(\to \gamma\gamma))$ [$\mathrm{fb}$] & 0.20 & 0.20 & 0.08 & 0.15 & 0.17 \\  
    %$\chi ^2$ & 85.9 & 89.3 & 93.7 & 87.3 & 86.6 \\  
    %P-value & 0.93 & 0.89 & 0.81 & 0.92 & 0.93 \\ 
    Br($B\to X_s\gamma$)$\ [\times 10^{-4}]$ & 3.15 & 3.11 & 3.03 & 3.09 & 3.11  \\  
    Br($B_d\to \mu ^+\mu ^-$)$\ [\times 10^{-10}]$ & 0.96 & 0.97 & 1.00 & 0.98 & 0.97  \\  
    Br($B_s\to \mu ^+\mu ^-$)$\ [\times 10^{-9}]$ & 2.99 & 3.04 & 3.14 & 3.07 & 3.14  \\  \bottomrule
\end{tabular}}
\label{t00}
\end{table*}

There is an important background of $ Z \to e^+e^- $ that should be considered, as it can have a relatively large cross section and the two electrons may be misidentified as photons.  
However, when associated with top-quark pair production, the cross section of $ pp \to t\bar{t}Z (\to e^+e^-) $ is significantly reduced, reaching a level comparable to that of the signal.  
Moreover, the probability of an electron being misidentified as a photon is typically below 0.1~\cite{CMS:2015myp}, indicating that this background is negligible in the present analysis. 
Similarly, jets can also be misidentified as photons, but the corresponding misidentification rate is usually very low, ranging from $ 10^{-3} $ to $ 10^{-5} $~\cite{Abdy:2024afj}.  
Therefore, these reducible backgrounds are not discussed further in this study.

The other SM backgrounds can be categorized into two types: resonant and non-resonant.
For the resonant backgrounds, we consider the processes $t\bar{t}H$ and $tjH$, where $H$ denotes the 125 GeV SM-like Higgs boson. These correspond to the following decay chains:  
\begin{equation}
\begin{aligned}
    &pp \to t(\to W^+ b)\, \bar{t}(\to W^- \bar{b})\, H(\to \gamma\gamma), \\ \quad &pp \to t(\to W^+ b)\, j\, H(\to \gamma\gamma),
\end{aligned}    
\end{equation}

respectively.
For the non-resonant backgrounds, we include the processes $t\bar{t}\gamma\gamma$,   
$b\bar{b}\gamma\gamma$, $tj\gamma\gamma$, $t\bar{t}\gamma$, and $Wjj\gamma\gamma$, corresponding to:

\begin{align}
&pp \to t(\to W^+ b)\, \bar{t}(\to W^- \bar{b})\, \gamma\gamma, \nonumber\\
&pp \to b\bar{b}\, \gamma\gamma, \nonumber\\
&pp \to t(\to W^+ b)\, j\, \gamma\gamma, \quad \\
&pp \to t(\to W^+ b)\, \bar{t}(\to W^- \bar{b})\, \gamma, \nonumber\\
&pp \to W(\to \ell \nu)\, j\, j\, \gamma\gamma. \nonumber
\end{align}

Monte Carlo simulations for both signal and background processes are performed using the \textsf{MadGraph5\_aMC@NLO\_v3.4.2} package~\cite{Alwall:2011uj, Alwall:2014hca}.  
In the collider simulation, parton distribution functions (PDFs) are set to \textsf{NNPDF23LO1} via \textsf{LHAPDF6}~\cite{NNPDF:2014otw, Buckley:2014ana}. 
Particle decays, parton showering, and hadronization are handled by \textsf{PYTHIA\_v8.2}~\cite{Sjostrand:2014zea}, interfaced through the \textsf{MG5a-MC\_PY8\_interface}. Jet clustering is performed using the \textsf{anti-$k_T$} algorithm~\cite{Cacciari:2008gp}.  
Detector effects are simulated with \textsf{Delphes-3.5.0},  using the official CMS card provided with the package~\cite{deFavereau:2013fsa, Selvaggi:2014mya}. The final analysis of the simulated events is carried out using \textsf{MadAnalysis5 v1.9.60}~\cite{Conte:2012fm}.

\begin{figure*}[!tbp] 
\centering 
\includegraphics[width=1\linewidth]{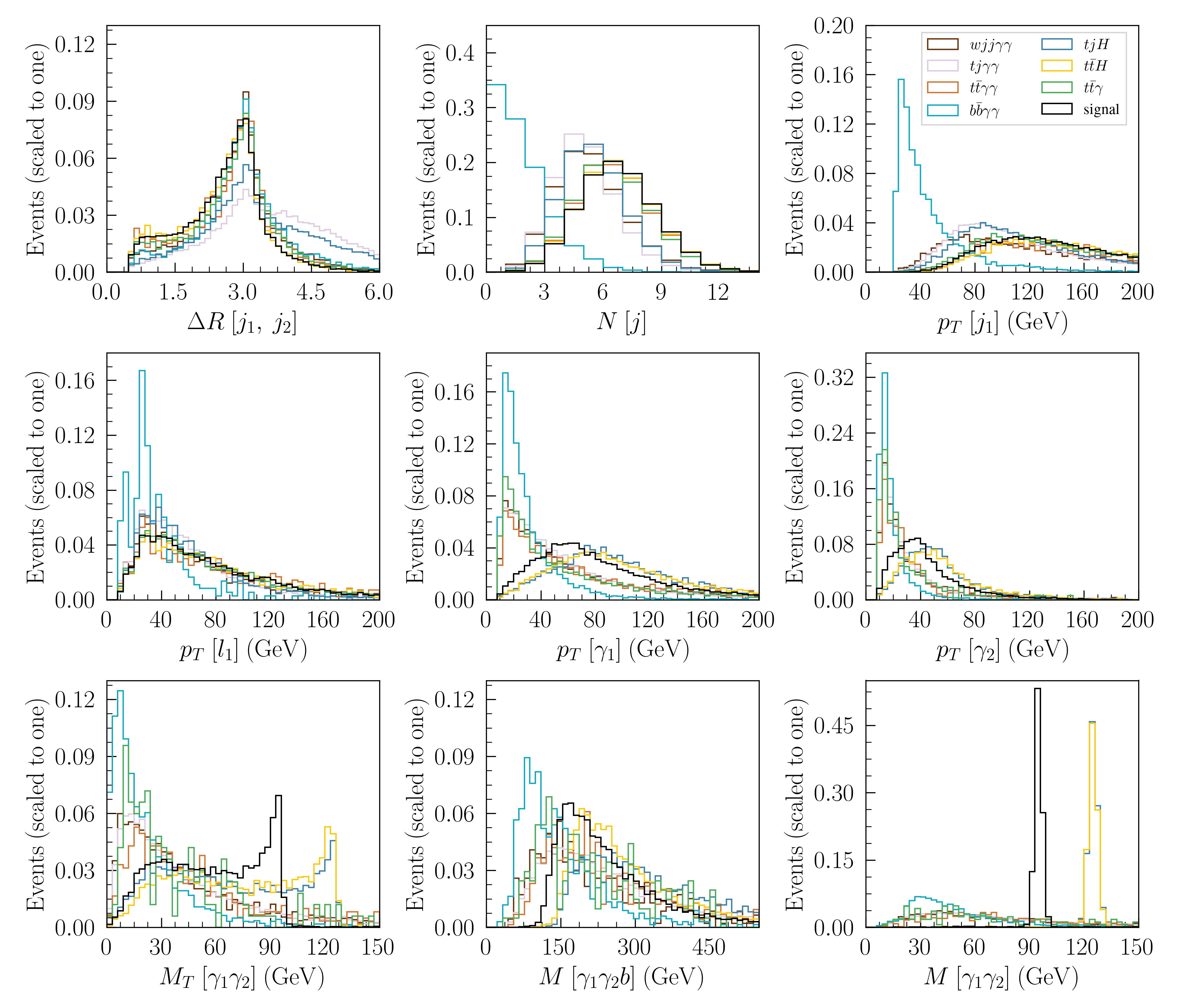}
\caption{\label{ma} 
The normalized distributions of $\Delta R[j_1,\ j_2]$ (upper left), N$[j]$ (upper middle), $p_T^{j_1}$(upper right), $p_T^{l_1}$ (upper middle), $p_T^{\gamma_1}$(middle), $p_T^{\gamma_2}$ (middle right), $M_T [\gamma_1\gamma_2]$ (lower left), $M[\gamma_1\gamma_2]$ (lower middle) and  $M[\gamma_1\gamma_2b]$ (lower right) for the signal and SM backgrounds at 14 TeV HL-LHC. } 
\end{figure*}

Monte Carlo simulations are performed at the 14 TeV HL-LHC, the 27 TeV HE-LHC~\cite{Cepeda:2019klc}, and the 100 TeV FCC-hh~\cite{FCC:2018vvp}, which are designed to deliver integrated luminosities of 3$\abm$, 10$\abm$, and 30$\abm$, respectively. 
In Fig.~\ref{ma}, the normalized distributions of several kinematic observables for the signal and SM backgrounds at the 14 TeV HL-LHC are shown, including:
$\Delta R[j_1,\ j_2]$ (upper left), the jet multiplicity N$[j]$ (upper middle), the transverse momenta of the leading jet $p_T^{j_1}$ (upper right), the leading lepton $p_T^{\ell_1}$ (middle left), the leading and subleading photons $p_T^{\gamma_1}$ (middle middle) and $p_T^{\gamma_2}$ (middle right), the transverse mass $M_T[\gamma_1\gamma_2]$ (lower left), the invariant mass $M[\gamma_1\gamma_2]$ (lower middle), and $M[\gamma_1\gamma_2b]$ (lower right).
In each case, $\gamma_1$ denotes the photon with the highest energy, $\gamma_2$ the next highest, and similarly for $j_1$ and $j_2$. 
Here, the invariant mass $M[\gamma_1\gamma_2]$ is defined as
\begin{widetext}
\begin{align}
    M[\gamma_1\gamma_2]\equiv \sqrt{(E^{\gamma_1}+E^{\gamma_2})^2 - (p^{\gamma_1}_1+p^{\gamma_2}_1)^2- (p^{\gamma_1}_2+p^{\gamma_2}_2)^2-(p^{\gamma_1}_3+p^{\gamma_2}_3)^2} \, ,
\end{align}
\end{widetext}
and the transverse masses $M_T[\gamma_1\gamma_2]$ is defined as
\begin{align}
     M_T[\gamma_1\gamma_2]\equiv \sqrt{(E^{\gamma_1}+E^{\gamma_2})^2 - (p^{\gamma_1}_1+p^{\gamma_2}_1)^2- (p^{\gamma_1}_2+p^{\gamma_2}_2)^2}.
\end{align}
The definition of the invariant mass $M[\gamma_1\gamma_2b]$ follows the same form as that of $M[\gamma_1\gamma_2]$, with the inclusion of the four-momentum of the $b$-jet.
From this figure, one can notice the following details:
\begin{itemize}

    \item From the upper left panel of Fig.~\ref{ma}, one can see that the backgrounds of $tjH$ and $tj\gamma\gamma$ have more events clustered at larger $\Delta R[j_1,\ j_2]$ values compared to the signal. Thus, choosing an appropriate $\Delta R[j_1,\ j_2]$ can effectively reduce these backgrounds.

    \item The upper middle panel of Fig.~\ref{ma} shows that the $b\bar{b}\gamma\gamma$ background typically has fewer jets than others. By choosing a higher jet count threshold, $N[j]$, one can significantly reduce this background.

    \item The upper right, middle, and middle right panels of Fig.~\ref{ma} show that the non-resonant backgrounds, especially $b\bar{b}\gamma\gamma$, typically have lower transverse momenta for $j_1$, $\gamma_1$, and $\gamma_2$. Consequently, choosing higher transverse momentum thresholds for these particles can effectively suppress the non-resonant background. 

    \item From the lower left panel of Fig.~\ref{ma}, one can observe that the transverse mass $M_T[\gamma_1\gamma_2]$ for non-resonant backgrounds tends to be distributed at lower values. In contrast, the distributions for the signal and resonant backgrounds peak at higher values and exhibit sharp cutoffs near their respective resonance masses, i.e., 95 GeV for the signal and 125 GeV for the SM-like Higgs.

    \item From the lower middle panel of Fig.~\ref{ma}, one can observe that both the resonant background and the signal show a preference for a larger invariant mass of $\gamma_1$, $\gamma_2$, and $b$. Consequently, choosing a higher $M[\gamma_1\gamma_2b]$ can effectively reduce the non-resonant background.

    \item From the lower middle panel of Fig.~\ref{ma}, one can observe that the invariant mass $M[\gamma_1\gamma_2]$ distributions for the signal and resonant backgrounds exhibit narrow peaks centered around their respective resonance masses, i.e., 95 GeV for the signal and 125 GeV for the SM-like Higgs.
    In contrast, the non-resonant backgrounds show broader distributions with lower invariant mass values and lack any distinct peak structure.

\end{itemize}

\subsection{\label{Cuts}Events selecte and Cut flow}

The events selected for analysis are required to satisfy the following basic kinematic criteria for leptons, jets, and photons:
\begin{equation}
\begin{aligned}
p_{T}^{j/b}>20\ \mathrm{GeV},\ p_{T}^{\ell/\gamma}&>10\ \mathrm{GeV},\ |\eta_{i}|<5\,,\\
\Delta R[i,\ j]>0.4\,&.\ (i,j=j,b,\ell,\gamma)
\end{aligned}
\end{equation}

In addition, at least one $W$ boson from the top quark decays is required to decay leptonically (i.e., into a lepton and a neutrino), which helps suppress the $b\bar{b}\gamma\gamma$ background.
Furthermore, two isolated photons are required in the final state to allow for invariant mass reconstruction.
Based on the distributions shown in Fig.~\ref{ma}, the following selection cuts are applied in the Monte Carlo simulations for the HL-LHC:
\begin{itemize}

    \item $\textbf{Basic Cut}$: We select events containing at least one lepton ($N(\ell) \geq 1$), two photons ($N(\gamma) \geq 2$), two jets ($N(j) \geq 2$), and at least one b-jet ($N(b) \geq 1$). For the transverse momentum ($p_T$) of jets and photons, the following requirements apply:
    \begin{equation}
    \begin{aligned}
    p_T[\gamma_1] > 30\ \mathrm{GeV}&,\ p_T[j_1] > 40\ \mathrm{GeV} , \\  p_T[j_2]  &< 220\ \mathrm{GeV}.
    \end{aligned}    
    \end{equation}
    
    The spatial separation between jets and photons should satisfy the following criteria:
    \begin{align}
    \Delta R[j_1,\ j_2] < 4\, , \ \Delta R[\gamma_1,\ \gamma_2] < 3.6\,.
    \end{align}
    To further suppress the background from $t\bar{t}\gamma$, we require that the phase angle of the b-jet ($\theta_{b_1}$), the pseudorapidity of the jet and lepton ($\eta_{j_1}$ and $\eta_{\ell_1}$), and the Lorentz factor of the jet ($\gamma_{j_1}$) meet the following conditions:
    \begin{equation}
    \begin{aligned}
    0.2 < \theta_{b_1} < 3\,,\ \eta_{j_1} > -2.6\, \\ \eta_{\ell_1} > -2.5 ,\  \gamma_{j_1} < 50\,.
    \end{aligned}        
    \end{equation}

    \item $\textbf{Mass  Cut}$: For the invariant masses of the diphoton pair ($M[\gamma_1\gamma_2]$) and the photon with jet ($M[\gamma_2j_1]$), we choose:
    \begin{equation}
    \begin{aligned}
    92\  \mathrm{GeV} < M[\gamma_1\gamma_2] < 98&\  \mathrm{GeV},\\   M[\gamma_2j_1] <390 \  \mathrm{GeV}.
    \end{aligned}        
    \end{equation}

     For the transverse masses of the photon, diphoton, and diphoton with a b-jet ($M_T[\gamma_1]$, $M_T[\gamma_1\gamma_2]$, and $M_T[\gamma_1\gamma_2b_1]$), we select:
\begin{equation}
\begin{aligned}
    M_T[\gamma_1] <300\  \mathrm{GeV},\ 
    M_T[\gamma_1\gamma_2] <&\ 96\  \mathrm{GeV},\\  
    M_T[\gamma_1\gamma_2b_1] <440\  \mathrm{GeV}.  
\end{aligned}
\end{equation}

    \item $\textbf{Energy  Cut}$: The transverse energy ($E_T$), transverse hadronic energy ($H_T$), missing transverse energy ($\SLASH{E}{.2}_T$), and missing transverse hadronic energy ($\SLASH{H}{.2}_T$) should meet the following criteria:
\begin{equation}
   \begin{aligned}
    E_T >200 \ \mathrm{GeV}, \ H_T <900 \ \mathrm{GeV}, \\ 
    10\ \mathrm{GeV} < \SLASH{E}{.2}_T < 190\  \mathrm{GeV} , \  
    \SLASH{H}{.2}_T >30 \ \mathrm{GeV}  . 
    \end{aligned} 
\end{equation}

    \end{itemize}

\begin{table*}[!tbp]
\centering
\caption{Cut flow of cross sections at the 14 TeV HL-LHC for benchmark point P1 with an integrated luminosity of $L = 300  \fbm$, as detailed in Table~\ref{t00}.}
\setlength{\tabcolsep}{13pt}
\resizebox*{1\textwidth}{!}{
\begin{tabular}{@{\hspace{20pt}}ccccccccc@{\hspace{20pt}}}
    \toprule
    %    &Signal&\multicolumn{7}{c}{Backgrounds}  \\ \midrule
    % Cuts & signal & $tt\gamma$ & $tth$ & $tjh$ & $bb\gamma\gamma$ & $tt\gamma\gamma$ & $tj\gamma\gamma$ & $Wjj\gamma\gamma$ \\ \midrule
    \multirow{2}{*}{Cuts} &  Signal$(\sigma \times L)$  & \multicolumn{7}{c}{Background$(\sigma\times L )$}                             \\ \cmidrule(r){2-2} % 仅为signal列绘制线条
    \cmidrule(l){3-9} % 为background下的所有列绘制线条
    & $t\bar{t}h$ & $t\bar{t}\gamma$ & $t\bar{t}H$ & $tjH$ & $b\bar{b}\gamma\gamma$ & $t\bar{t}\gamma\gamma$ & $tj\gamma\gamma$ & $Wjj\gamma\gamma$ \\ \midrule
    Initial & 60 & 1536000 & 168 & 23 & 2293800 & 3354 & 5253 & 65400 \\ 
    Basic cut & 5.27 & 4577.3 & 17.25 & 0.712 & 848.7 & 200.0 & 89.61 & 268.1 \\ 
    Mass cut & 4.13 & 46.08 & 0.029 & 0.0 & 0.0 & 6.47 & 2.67 & 4.58 \\ 
    Energy cut & 3.76 & 0.0 & 0.029 & 0.0 & 0.0 & 5.8 & 1.91 & 3.27 \\ \bottomrule
\end{tabular}}
\label{t01}		
\end{table*}

These cuts are performed on both our reference points and background events to calculate the signal significance. The results of these sequential cuts are detailed in Table~\ref{t01}, where the integrated luminosity is set to 300 $\text{fb}^{-1}$. The dominant backgrounds, $Wjj\gamma\gamma$, $b\bar{b}\gamma\gamma$, and $t\bar{t}\gamma$, are significantly reduced by more than two to three orders of magnitude after the base cut, while the signal reduction is limited to one order of magnitude. Subsequent cuts at the invariant mass and transverse mass effectively preserved the signal while further reducing the background. After these cuts, $t\bar{t}\gamma$, $t\bar{t}\gamma\gamma$, $tj\gamma\gamma$, and $Wjj\gamma\gamma$ showed reductions of more than an order of magnitude, while $t\bar{t}H$, $tjH$, and $b\bar{b}\gamma\gamma$ were nearly eliminated. The final cut on transverse energy and transverse hadronic energy additionally minimized the $t\bar{t}\gamma$ background to negligible levels. In the end, only minimal instances of $t\bar{t}H$, $t\bar{t}\gamma\gamma$, $tj\gamma\gamma$, and $Wjj\gamma\gamma$ remained.

\begin{table*}[!tbp]
\centering
\caption{Cut flow of the cross sections at 27 TeV HE-LHC for benchmark point P1 with an integrated luminosity of $L = 300  \fbm$, as detailed in Table~\ref{t00}.}
\setlength{\tabcolsep}{15pt}
\resizebox*{1\textwidth}{!}{
\begin{tabular}{@{\hspace{15pt}}ccccccccc@{\hspace{15pt}}}
    \toprule
    % &Signal&\multicolumn{7}{c}{Backgrounds}  \\ \midrule
    % Cuts & signal & $tt\gamma$ & $tth$ & $tjh$ & $bb\gamma\gamma$ & $tt\gamma\gamma$ & $tj\gamma\gamma$ & $Wjj\gamma\gamma$ \\ \midrule
        \multirow{2}{*}{Cuts} &  Signal$(\sigma \times L)$  & \multicolumn{7}{c}{Background$(\sigma\times L )$}                             \\ \cmidrule(r){2-2} % 仅为signal列绘制线条
    \cmidrule(l){3-9} % 为background下的所有列绘制线条
    & $t\bar{t}h$ & $t\bar{t}\gamma$ & $t\bar{t}H$ & $tjH$ & $b\bar{b}\gamma\gamma$ & $t\bar{t}\gamma\gamma$ & $tj\gamma\gamma$ & $Wjj\gamma\gamma$ \\ \midrule
    Initial & 192 & 3168000 & 927 & 115 & 3408000 & 14178 & 19920 & 206009 \\ 
    Basic cut & 13.92 & 6874.6 & 80.06 & 3.23 & 1192.8 & 776.6 & 358.5 & 995.0 \\ 
     Mass cut & 9.14 & 0.0 & 0.111 & 0.002 & 68.16 & 17.22 & 5.13 & 12.36 \\ 
    Energy cut & 8.59 & 0.0 & 0.111 & 0.002 & 0.0 & 16.27 & 3.21 & 10.3 \\ \bottomrule
\end{tabular}}
\label{t02}		
\end{table*}

\begin{table*}[!tbp]
    \centering
    \caption{Cut flow of the cross sections at 100 TeV FCC-hh for benchmark point P1 with an integrated luminosity of $L = 300  \fbm$, as detailed in Table~\ref{t00}.}	
    \setlength{\tabcolsep}{15pt}
    \resizebox*{1\textwidth}{!}{
    \begin{tabular}{@{\hspace{15pt}}ccccccccc@{\hspace{15pt}}}
        \toprule
        % &Signal&\multicolumn{7}{c}{Backgrounds}  \\ \midrule
        % Cuts & signal & $tt\gamma$ & $tth$ & $tjh$ & $bb\gamma\gamma$ & $tt\gamma\gamma$ & $tj\gamma\gamma$ & $Wjj\gamma\gamma$ \\ \midrule
        \multirow{2}{*}{Cuts} &  Signal$(\sigma \times L )$  & \multicolumn{7}{c}{Background$(\sigma\times L )$}                             \\ \cmidrule(r){2-2} % 仅为signal列绘制线条
    \cmidrule(l){3-9} % 为background下的所有列绘制线条
    & $t\bar{t}h$ & $t\bar{t}\gamma$ & $t\bar{t}H$ & $tjH$ & $b\bar{b}\gamma\gamma$ & $t\bar{t}\gamma\gamma$ & $tj\gamma\gamma$ & $Wjj\gamma\gamma$ \\ \midrule
        Initial & 2325 & 33120000 & 11724 & 1005 & 14856000 & 141570 & 138120 & 1255800 \\ 
        Basic cut &101.42 & 53985 & 621 & 17.1 & 6388.1 & 4764.2 & 1446.7 & 4897.6 \\ 
         Mass cut & 64.94 & 662.4 & 1.06 & 0.0 & 148.6 & 91.0 & 24.82 & 37.67 \\ 
        Energy cut & 59.75 & 0.0 & 0.821 & 0.0 & 0.0 & 78.86 & 21.28 & 25.12 \\ \bottomrule		
    \end{tabular}}
    \label{t03}		
\end{table*}

The distributions of various kinematic variables for the signal and SM backgrounds at the 27 TeV HE-LHC and the 100 TeV FCC-hh have been examined and found to be similar to those at the HL-LHC, as shown in Fig.~\ref{ma}.
In contrast to the Monte Carlo simulations for the HL-LHC, the following selection cuts are applied in the simulations for the HE-LHC:
\begin{itemize}

    \item $\textbf{Basic Cut}$: We select events that include at least one lepton ($N(\ell) \geq 1$), two photons ($N(\gamma) \geq 2$), two jets ($N(j) \geq 2$), and at least one b-jet ($N(b) \geq 1$). For the transverse momentum ($p_T$) of jets, leptons, and photons, the following cuts are applied:
    \begin{equation}
    \begin{aligned}
     p_T[j_1] >30 \  \mathrm{GeV},\  
     p_T[\ell_1] >&\ 20\  \mathrm{GeV},\\ 
     p_T[\gamma_1]  >30\ \mathrm{GeV}.
    \end{aligned}
    \end{equation}

    The spatial separation between jets and photons should meet the following criteria: 
    \begin{align}
    \Delta R[j_1, j_2] <4\,,\ 
    \Delta R[\gamma_1, \gamma_2] <3.4\,.
    \end{align}
    To further reduce the background from $t\bar{t}\gamma$, the pseudorapidity of the jet and lepton ($\eta_{j_1}$ and $\eta_{\ell_1}$) should meet the following criteria:
    \begin{align}
    \eta_{j_1} >-3.4\,, \ 
    \eta_{\ell_1} <3.5\,.
    \end{align}

    \item $\textbf{Mass Cut}$: For the invariant masses of the diphoton pair ($M[\gamma_1\gamma_2]$) and the diphoton with a b-jet ($M[\gamma_1\gamma_2b_1]$), we require: 
    \begin{equation}
    \begin{aligned}
    93\  \mathrm{GeV} < M[\gamma_1\gamma_2]< 97\  \mathrm{GeV}, \\ 
    M[\gamma_1\gamma_2b_1]> 120 \ \mathrm{GeV}.
    \end{aligned}
    \end{equation}

    For the transverse masses of the lepton, photon, diphoton, and diphoton with a b-jet ($M_T[\ell_1]$, $M_T[\gamma_1]$, $M_T[\gamma_1\gamma_2b]$, and $M_T[\gamma_1\gamma_2b_1]$), we require:
\begin{equation}
\begin{aligned}
    M_T[\ell_1] <40\ \mathrm{GeV},\ 
    M_T[\gamma_1] <380\ \mathrm{GeV},\\ 
    M_T[\gamma_1\gamma_2b] >6\ \mathrm{GeV},\ M_T[\gamma_1\gamma_2b_1] <500\ \mathrm{GeV}  .
\end{aligned}
\end{equation}

    \item $\textbf{Energy Cut}$: The missing transverse energy ($\SLASH{E}{.2}_T$) is used to reduce the background of $Wjj\gamma\gamma$, which require
    \begin{equation}
     \SLASH{E}{.2}_T > 15\  \mathrm{GeV}. 
    \end{equation}	
    
\end{itemize}

The cut flow for signal and background at the HE-LHC with an integrated luminosity of $L=300\ \text{fb}^{-1}$ is shown in Table~\ref{t02}. The backgrounds of $t\bar{t}\gamma$, $b\bar{b}\gamma\gamma$ and $Wjj\gamma\gamma$ still play an important role. After the basic cut, the backgrounds of $t\bar{t}\gamma$ and $b\bar{b}\gamma\gamma$ are reduced by over two and three orders of magnitude, respectively. These backgrounds are further suppressed by subsequent cuts at the invariant mass and transverse mass. The $t\bar{t}\gamma$ background is nearly eliminated and no longer dominant. The $t\bar{t}H$ and $tjH$ backgrounds become nearly negligible. Finally, the missing transverse energy cut effectively eliminates the $b\bar{b}\gamma\gamma$ background. However, the $t\bar{t}\gamma\gamma$, $tj\gamma\gamma$, and $Wjj\gamma\gamma$ backgrounds are difficult to remove and ultimately persist.

The following selection cuts are applied in the FCC-hh simulations:
\begin{itemize}

    \item $\textbf{Basic Cut}$: We select events containing at least one lepton ($N(\ell) \geq 1$), two photons ($N(\gamma) \geq 2$), two jets ($N(j) \geq 2$), and at least one b-jet ($N(b) \geq 1$). For the transverse momentum ($p_T$) of the jets and photons, we have the following requirements
    \begin{equation}
    \begin{aligned}
     p_T[\gamma_1] >\ 35 \  \mathrm{GeV},\ 
     p_T[j_2] <\ &220\  \mathrm{GeV},\\ 
     p_T[b_1]  >\ 25\ \mathrm{GeV}.
    \end{aligned}  
    \end{equation}

    The spatial separation between jets and photons should meet the following criteria:
    \begin{align}
    \Delta R[j_1, j_2] <\ 4\,,\ \Delta R[\gamma_1, \gamma_2] <\ 3.4\,.
    \end{align}
    To further reduce the background from $t\bar{t}\gamma$, we require that the phase angle of the lepton ($\theta_{\ell_1}$) and the pseudorapidity of the jet ($\eta_{j_1}$) meet the following criteria:
    \begin{align}
    \theta_{\ell_1} > 0.1\,,\ \eta_{j_1} > -3.8\,. 
    \end{align}

    \item $\textbf{Mass Cut}$: For the invariant mass of the diphoton pair $M[\gamma_1\gamma_2]$, we choose the following criteria:
    \begin{align}
    93\  \mathrm{GeV} < M[\gamma_1\gamma_2] < 97\  \mathrm{GeV}.
    \end{align}
    For the transverse mass of the photon and lepton, we apply the following criteria:
    \begin{align}
    M_T[\gamma_1] < 330\  \mathrm{GeV}\ , \ M_T[\ell_1] < 295\  \mathrm{GeV}.
    \end{align}

    \item $\textbf{Energy Cut}$: The transverse hadronic energy ($H_T$) and the missing transverse energy ($\SLASH{E}{.2}_T$) should meet the following criteria:
    \begin{align}
      150\ \mathrm{GeV} < H_T <\ 900 \ \mathrm{GeV}, \  \SLASH{E}{.2}_T >\ 10\  \mathrm{GeV}. 
    \end{align}	
    
\end{itemize}

The cut flow of the cross sections for signal and background at the FCC-hh with an integrated luminosity of $L = 300 \ \text{fb}^{-1}$ is shown in Table~\ref{t03}. As the collision energy increases from 27 to 100 TeV, the cross sections for all signals and backgrounds, except for the $b\bar{b}\gamma\gamma$ background, increase about ten times. The primary backgrounds are $t\bar{t}\gamma$, $b\bar{b}\gamma\gamma$, and $Wjj\gamma\gamma$. The basic cut significantly reduces these backgrounds to a rare occurrence. After the basic cut, the backgrounds $b\bar{b}\gamma\gamma$, $t\bar{t}\gamma\gamma$, $tj\gamma\gamma$, and $Wjj\gamma\gamma$ become equally significant, each retaining a few thousand events. The resonance background $tjH$ is almost eliminated by the diphoton invariant mass cut, and $t\bar{t}H$ is significantly reduced by the same cut. After the energy cut, both $t\bar{t}\gamma$ and $b\bar{b}\gamma\gamma$ are nearly eliminated. Finally, the remaining backgrounds include $t\bar{t}\gamma\gamma$, $tj\gamma\gamma$, $Wjj\gamma\gamma$, and a small amount of $t\bar{t}H$.

%\end{itemize}

\subsection{\label{DsC}The discussion on future colliders}

In the analysis of the Monte Carlo simulation results, the statistical significance is evaluated using the Poisson formula \cite{Cowan:2010js}:
\begin{align}
\mathcal{S} = \sqrt{2L[(S+B)\mathrm{ln}(1+S/B) - S]} ,
\end{align}
where $S$ and $B$ are cross sections of signal and background, respectively, while $L$ represents the integrated luminosity.  
With an integrated luminosity of $L = 300\fbm$, the statistical significance of the benchmark point P1 from Table~\ref{t00} can reach $1.1\,\sigma$ at the HL-LHC, $1.5\,\sigma$ at the HE-LHC, and $5.0\,\sigma$ at the FCC-hh, respectively.
It is evident that, for the same integrated luminosity, a higher collider energy leads to a greater statistical significance.

\begin{figure*}[!tbp] 
\centering 
\includegraphics[width=\linewidth]{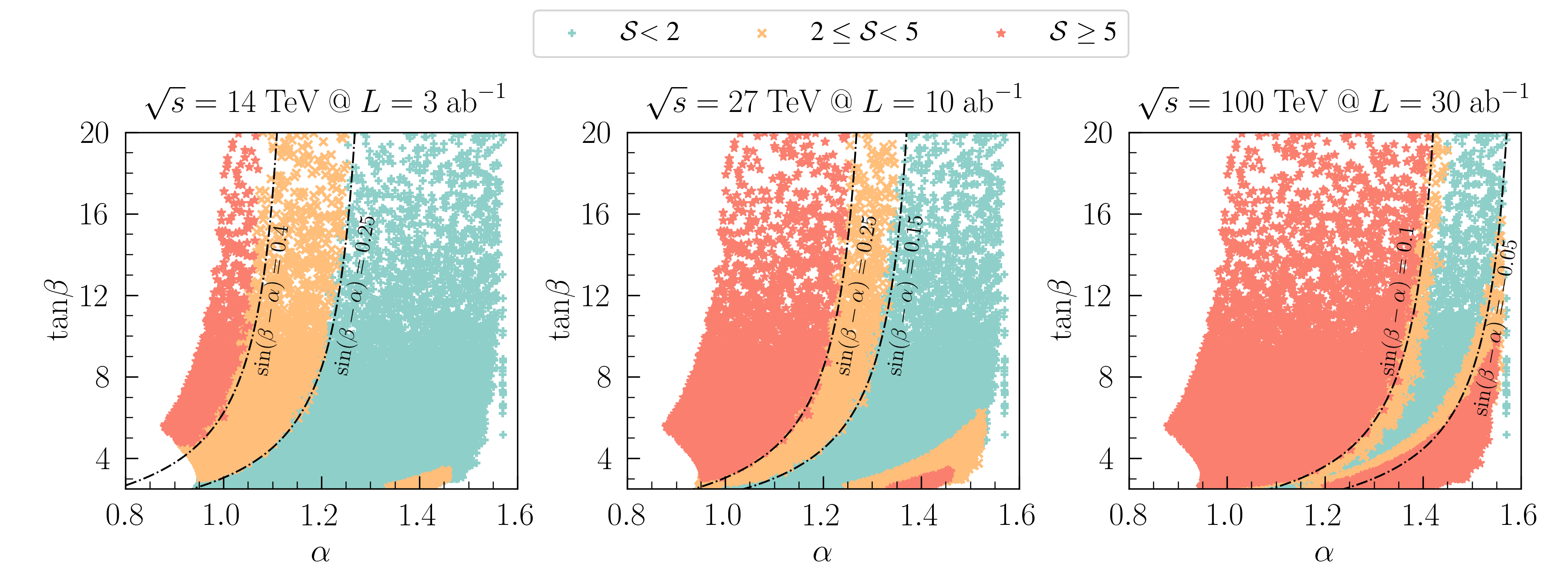}
\caption{\label{sig} 
    Signal statistical significance of surviving samples in the $\tan\beta$ versus $\alpha$ planes at the 14 TeV HL-LHC for integrated luminosities of $L = 3 \ \abm$ (left), the 27 TeV HE-LHC for integrated luminosities of $L = 10 \ \abm$ (middle), and the 100 TeV FCC-hh for $L = 30 \ \text{ab}^{-1}$ (right).}
\end{figure*}

In Fig.~\ref{sig}, the signal statistical significance of the surviving samples is shown in the $\tan\beta$ versus $\alpha$ plane for the 14 TeV HL-LHC with an integrated luminosity of $L = 3\abm$ (left), the 27 TeV HE-LHC with $L = 10\abm$ (middle), and the 100 TeV FCC-hh with $L = 30\abm$ (right).

According to the MC simulation results, the minimum scattering cross-section required to achieve a signal significance of 5$\sigma$ at the HL-LHC with $L = 3\abm$ is 0.3 fb of the process $pp \to t\bar{t}(h \to \gamma\gamma)$. 
In contrast, to achieve the same significance with same $L$ at HE-LHC and FCC-hh, the required cross-sections are 0.67 fb and 2.36 fb, respectively. 
Meanwhile, in our parameter scan, the maximum cross section obtained for the surviving samples at the HL-LHC is approximately 0.44 fb and it require $L=240\fbm$ to be covered at the 2$\sigma$ level and $L = 1.4\abm$ to covered at 5$\sigma$ level.
For this sample, to reach a $2\sigma$ ($5\sigma$) signal significance at the HE-LHC and the FCC-hh require $L = 120 \fbm\ (750\fbm) $ and $L = 12\fbm\ (71\fbm)$, respectively.

With increasing collision energy and integrated luminosity, most of the surviving samples can be probed at the $2\sigma$ or even $5\sigma$ level at the 100 TeV FCC-hh with an integrated luminosity of $L = 30\abm$.
At the 14 TeV HL-LHC with an integrated luminosity of $3\,\abm$, parameter regions with $\sin(\beta-\alpha) \gtrsim 0.4$ and $\sin(\beta-\alpha) \gtrsim 0.25$ can be probed at the $5\sigma$ and $2\sigma$ significance levels, respectively. At the 27 TeV HE-LHC with $L = 10\,\abm$, the sensitivity improves to $\sin(\beta-\alpha) \gtrsim 0.25$ ($5\sigma$) and $\gtrsim 0.15$ ($2\sigma$). For the 100 TeV FCC-hh with $L = 30\,\abm$, even regions with $\sin(\beta-\alpha) \gtrsim 0.1$ or $\sin(\beta-\alpha) \lesssim -0.05$ can be covered at the $5\sigma$ level.

However, there remain some surviving samples with the alignment condition \ $\sin(\beta-\alpha)\approx0$ that are difficult to probe even at the $2\sigma$ level in diphoton channel, even further increases in energy and luminosity.
This is because these samples exhibit highly suppressed diphoton decay rates, making them challenging to detect in this channel. 
As a result, alternative search channels or complementary methods are required to test these scenarios.
In this region, the decays $h\to b\bar{b}$ or $h\to\tau^+\tau^-$ may become dominant. These channels could potentially complement the parameter space coverage unattainable through the diphoton channel, although they pose greater challenges due to their more complex backgrounds. The relevant study will be conducted in detail in our future work.

It is worth noting that if one focuses solely on the diphoton excess, the pseudoscalar particle $A$ could also serve as a viable candidate to explain this excess. In the 2HDM-I, the coupling of $A$ to the top quark is proportional to $\cot\beta$, so the cross section of $pp\to t\bar{t}A_{95}$ is strongly suppressed for large $\tan\beta$~\cite{Branco:2011iw}. Furthermore, the decay $A\to\gamma\gamma$, which lacks the $W$-boson loop contribution, may further suppress the $pp\to t\bar{t}A_{95}\to t\bar{t}\gamma\gamma$ rate. Consequently, this channel may display interesting phenomenology for small $\tan\beta$, while the large $\tan\beta$ region warrants further study.
In addition, the $t\bar{t}\gamma\gamma$ channel can also be used to probe other models beyond the 2HDM-I that feature new scalars coupling to top quarks and decaying into diphotons. If the scalar–top–top coupling is sufficiently strong and the branching ratio for the scalar decay into diphotons is not too small, this channel may offer promising sensitivity.

\section{\label{sec:conclusion}Conclusions}

In this study, we investigated the potential of the Type-I 2HDM to account for the 95 GeV diphoton excess observed at the LHC, under current theoretical and experimental constraints.
The impacts of various experimental data on the model parameters were analyzed in detail.
Our parameter space scan shows that the 2HDM-I can successfully accommodate the observed excess.
Monte Carlo simulations were then performed to assess the discovery potential of a 95 GeV Higgs boson at future hadron colliders. Particular attention was given to the process $pp \to t(\to W^+ b)\, \bar{t}(\to W^- \bar{b})\, h(\to \gamma\gamma)$, where the Higgs boson is produced in association with a top-quark pair and decays into diphotons.
The simulation results indicate that a large portion of the viable parameter space can be effectively probed at future colliders via the top-associated diphoton channel.

Based on our analysis, the following conclusions are drawn about the 95 GeV Higgs boson in the 2HDM-I:

\begin{itemize}

    \item The analysis reveals that direct Higgs search data impose strong constraints on the model parameter $\alpha$, excluding samples with $\alpha \lesssim 0.95$.Constraints from B-physics, particularly the branching ratio of $B \to X_s \gamma$, require $\tan\beta \gtrsim 2.6$. 
    %In addition, samples with $|m_{12}| \lesssim 13\, m_A + 500$ GeV can yield a $\Delta a_{\mu}$ consistent with the observed muon anomalous magnetic moment, potentially accounting for the discrepancy between theory and observation.

    \item The branching ratio of the lighter Higgs boson decaying into diphotons, $\text{Br}(h \to \gamma\gamma)$, primarily depends on the parameters $\alpha$ and $\tan\beta$. It reaches a maximum of approximately $10^{-1}$ with $\alpha\approx1.55$, and a minimum of about $10^{-6}$ around the alignment limit $\sin(\beta-\alpha)\approx0$.
    Meanwhile, both the production cross section $\sigma(gg \to h)$ and the reduced top-Higgs coupling $(C_{tth}/C_{tth}^{\rm SM})^2$ exhibit similar dependence on $\alpha$ and $\tan\beta$, being approximately proportional to $\cos^2\alpha$ and inversely proportional to $\sin^2\beta$.
    The 95 GeV diphoton excess observed by CMS and ATLAS can be explained within $1\sigma$ for a wide parameter range with $0.95\lesssim\alpha\lesssim1.55$.

    \item For the surviving samples, the maximum cross sections of the process $pp \to t\bar{t}(h \to \gamma\gamma)$ at the HL-LHC, HE-LHC, and FCC-hh can reach 0.44 fb, 1.42 fb, and 17.21 fb, respectively.
    The dominant SM backgrounds for this channel arise from the processes $t\bar{t}\gamma$, $b\bar{b}\gamma\gamma$, and $Wjj\gamma\gamma$.

    \item The Monte Carlo simulation results indicate that, for a fixed integrated luminosity, a higher collider energy leads to a greater statistical significance.
    According to the MC simulation results, the minimum scattering cross-section required to achieve a signal significance of 5$\sigma$ at the HL-LHC with $L = 3\abm$ is 0.3 fb of the process $pp \to t\bar{t}(h \to \gamma\gamma)$.
    In contrast, to achieve the same significance with the same $L$ at HE-LHC and FCC-hh, the required cross-sections are 0.67 fb and 2.36 fb, respectively.
    The required cross-sections for these processes are easily satisfied by the surviving samples. For these surviving samples, an integrated luminosity of at least $240\fbm$ ($1.4\abm$) is required to achieve a significance of $2\sigma$ ($5\sigma$) at the HL-LHC.
    In contrast, at the HE-LHC and FCC-hh, the required integrated luminosities are $120\fbm$ ($750\fbm$) for the HE-LHC and $12\fbm$ ($71\fbm$) for the FCC-hh, respectively.

    \item At the 14 TeV HL-LHC with an integrated luminosity of $3\,\abm$, parameter regions with $\sin(\beta-\alpha) \gtrsim 0.4$ and $\sin(\beta-\alpha) \gtrsim 0.25$ can be probed at the $5\sigma$ and $2\sigma$ significance levels, respectively. At the 27 TeV HE-LHC with $L = 10\,\abm$, the sensitivity improves to $\sin(\beta-\alpha) \gtrsim 0.25$ ($5\sigma$) and $\gtrsim 0.15$ ($2\sigma$). For the 100 TeV FCC-hh with $L = 30\,\abm$, even regions with $\sin(\beta-\alpha) \gtrsim 0.1$ or $\sin(\beta-\alpha) \lesssim -0.05$ can be covered at the $5\sigma$ level. 
    Samples located in the region $\sin(\beta-\alpha)\approx0$ are difficult to probe even at the $2\sigma$ level, despite further increases in collision energy and integrated luminosity.
    This is due to the suppression of the coupling $C_{hWW}$ in this region, which significantly reduces the diphoton decay rate of the light Higgs boson.
    Such scenarios may therefore need to be tested through alternative production or decay channels.

\end{itemize}

\acknowledgments
This work was supported by the National Natural Science Foundation of China under Grant No. 12275066 and by the startup research funds of Henan University. 
The work of K. Wang is also supported by the Open Project of the Shanghai Key Laboratory of Particle Physics and Cosmology under Grant No. 22DZ2229013-3.

% \appendix

% whether or not they are referenced in the text. This is an   appropriate
% \nocite{*}

% \bibliographystyle{apsrev4-2}
\bibliographystyle{apsrev4-1}
\bibliography{apssamp}% Produces the bibliography via BibTeX.

\end{document}